\documentclass{article}

\usepackage{microtype}
\usepackage{graphicx}
\usepackage{amsmath,amsfonts}
\usepackage{algorithmic}
\usepackage{algorithm}
\usepackage{array}
\usepackage{multirow}
\usepackage{makecell}
\usepackage{subcaption}
\usepackage{booktabs}

\usepackage{hyperref}

\usepackage[preprint]{icml2026}

\usepackage{amsmath}
\usepackage{amssymb}
\usepackage{mathtools}
\usepackage{amsthm}
\usepackage{bm}

\usepackage[capitalize,noabbrev]{cleveref}

\theoremstyle{plain}

\theoremstyle{definition}

\theoremstyle{remark}

\usepackage{xcolor}
\usepackage{color}
\usepackage{tabularx}
\usepackage[most]{tcolorbox}
\usepackage[textsize=tiny]{todonotes}

\icmltitlerunning{A Unified Audio Language Model with Text-Aligned Factorized Audio Tokenization}

\begin{document}

\twocolumn[
  \icmltitle{UniAudio 2.0: A Unified Audio Language Model with Text-Aligned Factorized Audio Tokenization}



  \icmlsetsymbol{equal}{*}

  \begin{icmlauthorlist}
    \icmlauthor{Dongchao Yang}{cuhk}
    \icmlauthor{Yuanyuan Wang}{cuhk}
    \icmlauthor{Dading Chong}{ic}
    \icmlauthor{Songxiang Liu}{ic}
    \icmlauthor{Xixin Wu}{cuhk}
    \icmlauthor{Helen Meng}{cuhk}
  \end{icmlauthorlist}

  \icmlaffiliation{cuhk}{The Chinese University of Hong Kong, China}
  \icmlaffiliation{ic}{Independent Researcher, China}

  \icmlcorrespondingauthor{Helen Meng}{hmmeng@se.cuhk.edu.hk}

  \icmlkeywords{Machine Learning, ICML}

  \vskip 0.3in
]



\printAffiliationsAndNotice{}  

\begin{abstract}
We study two foundational problems in audio language models: (1) how to design an audio tokenizer that can serve as an intermediate representation for both understanding and generation; and (2) how to build an audio foundation model that generalizes in few-shot and zero-shot settings, analogous to large language models.
To this end, we make the following two contributions.
First, we propose ReasoningCodec, a discrete audio codec that factorizes audio into (i) reasoning tokens, which encode text-aligned, high-level analysis and planning representations for audio understanding and hierarchical generation, and (ii) reconstruction tokens, which encode semantic-rich acoustic cues for high-fidelity waveform reconstruction. 
This design achieves understanding performance comparable to strong continuous representations while improving generation quality and reconstruction fidelity over prior discrete tokenizers.
Second, we introduce a unified autoregressive architecture for text and audio, together with multi-stage training and multi-task data construction. Using this framework, we train UniAudio 2.0 on 100B text tokens and 60B audio tokens.
Across a wide range of speech, sound, and music tasks, UniAudio 2.0 performs competitively on in-domain evaluations and demonstrates strong few-shot and zero-shot generalization to unseen tasks. Demo, code, and checkpoints will be available at \href{https://dongchaoyang.top/UniAudio2Demo/}{https://dongchaoyang.top/UniAudio2Demo/}.
\end{abstract}

\section{Introduction}
\label{introduction}
Large language models (LLMs) \cite{gpt4,llama3} have demonstrated remarkable success by unifying diverse language tasks under a single autoregressive framework. Inspired by this paradigm, recent research has applied similar modeling principles to the audio domain, such as LM-based audio generation tasks \cite{borsos2023audiolm,valle,speartts}, LM-based audio understanding tasks \cite{qwen2-audio,salmonn}, cross-modal interaction \cite{moshi,kimi-audio}. Despite rapid progress, however, current audio language models still fall short of the generalization, scalability, and task versatility exhibited by their text counterparts.
We argue that this limitation primarily stems from three fundamental challenges: the design of audio representations, the architecture of unified autoregressive models and the construction of large-scale multi-task training data.

\begin{figure*}[t]
    \centering
    \includegraphics[width=0.9\textwidth]{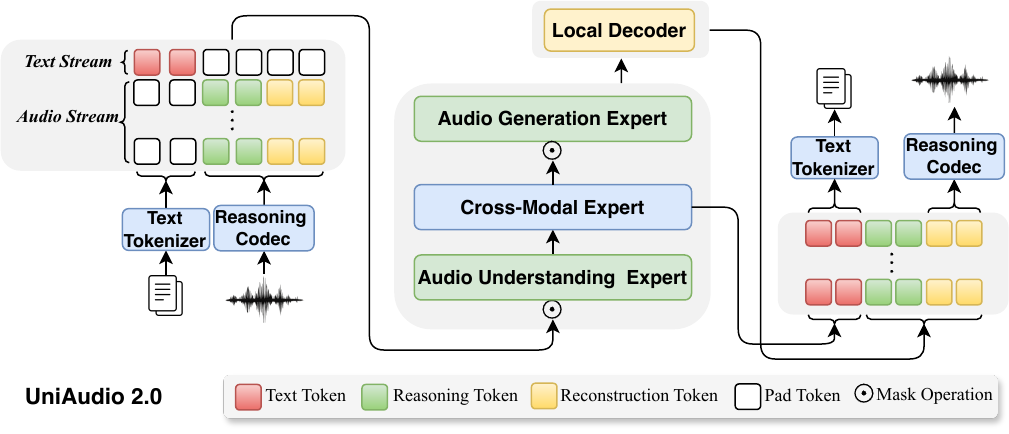}
    \caption{Overview of the proposed UniAudio~2.0 framework.
UniAudio~2.0 adopts a unified autoregressive architecture over text, reasoning, and reconstruction tokens, where reasoning tokens capture high-level, text-grounded semantics and reconstruction tokens preserve fine-grained acoustic details. The model integrates audio understanding, cross-modal, and audio generation experts to support unified audio understanding and generation.
}
    \label{fig:overview-uniaudio2}
\end{figure*}

On the representation side, existing approaches largely fall into two categories. 
Continuous representations, such as self-supervised representations (SSL features)~\cite{hubert,whisper}, are effective for perception and understanding tasks but are difficult to integrate into autoregressive audio generation due to the difficulty of modeling high-dimensional features. 
In contrast, discrete audio codecs~\cite{soundstream,encodec,dac,yang2023hifi,uniaudio1.5} enable efficient generation and scalable modeling, yet their tokens mainly encode low-level acoustic details and lack text-aligned, high-level abstractions for understanding. In this study, we focus on discrete tokenizers due to their scalability and compatibility with unified text-audio modeling objectives. 
To address their limited abstraction capability, we introduce ReasoningCodec, a novel audio codec that explicitly factorizes audio representations into reasoning tokens and reconstruction tokens. 
Reasoning tokens encode text-aligned, high-level analysis and planning representations that support audio understanding and hierarchical generation, while reconstruction tokens preserve semantic content and fine-grained acoustics for high-fidelity waveform reconstruction.

On the architectural side, most existing audio language models adopt a naive unified autoregressive transformer \cite{glm4-voice,moshi,kimi-audio} inherited from text LLMs, in which all layers indiscriminately process both text and audio tokens. Although such a design is simple and convenient, we argue that it is suboptimal for audio foundation models even with improved tokenization, because: (1) discrete audio tokens remain lossy, and propagating them uniformly across all layers can limit perceptual abstraction and reasoning for audio understanding; and (2) directly aligning text and audio tokens throughout all transformer layers is highly challenging and can lead to rapid forgetting of pre-trained textual knowledge. To address these challenges, we propose a unified autoregressive architecture with \textbf{functional layer specialization}. Rather than treating all transformer layers uniformly, we conceptually partition the model into three stages: the lower layers act as \textbf{audio understanding experts} that focus on perceptual abstraction and reasoning over audio; the intermediate layers serve as \textbf{cross-modal experts} to align and integrate text and audio, initialized from a pre-trained LLM (e.g., LLaMA3.2 3B) to preserve rich textual knowledge; and the upper layers act as \textbf{audio generation experts} that specialize in modeling fine-grained acoustics. This design maintains specialized inductive biases for understanding and generation while operating within a unified autoregressive framework.


On the data side, we curate large-scale open-sourced audio corpora spanning speech, sound, and music, and unify them into a diverse set of audio-centric tasks covering both understanding and generation.
Furthermore, inspired by sequential training in LVMs \cite{bai2023sequential}, we introduce the concept of \emph{auditory sentences}: long-context sequences composed of multiple segments that are linked by semantic or acoustic relations, where each segment can be an audio span, a text span (e.g., caption), or their paired form. Auditory sentences essentially serve as a \textit{unified task constructor}. By organizing multiple related segments into a single long-context sequence, an auditory sentence naturally induces a variety of task forms, including within-segment modeling (e.g., ASR/captioning), cross-segment dependency tracking (e.g., style/event consistency), and multi-step conditional generation (e.g., continuation conditioned on earlier segments).
This enables scalable multi-task pre-training without manually designing separate task-specific pipelines, while encouraging the model to reason over compositional structure and long-range dependencies.

Building on these design choices and the proposed multi-task data construction strategy, we train a unified audio understanding and generation model on 100B text tokens and 60B audio tokens, which we name UniAudio 2.0. Extensive experiments show that UniAudio 2.0 achieves competitive performance on seen tasks. Moreover, UniAudio 2.0 demonstrates strong few-shot and zero-shot generalization on a wide range of unseen tasks, highlighting its potential as a foundation model for audio language processing.

Our main contributions include:
\begin{itemize}
    \item \textbf{ReasoningCodec:} We propose a discrete audio tokenizer that factorizes audio into \emph{reasoning tokens} and \emph{reconstruction tokens}, enabling text-aligned high-level abstraction while preserving high-fidelity waveform reconstruction.
    \item \textbf{Functional layer specialization:} We introduce a unified autoregressive architecture that specializes lower, middle, and upper transformer layers into audio understanding, cross-modal alignment (initialized from a pretrained LLM), and audio generation experts, improving both cross-modal alignment and acoustic modeling.
    \item \textbf{Large-scale training and evaluation:} We curate a diverse set of audio-related tasks and introduce \emph{auditory sentences} for scalable multi-task pre-training. We then train {UniAudio 2.0} on 100B text tokens and 60B audio tokens across text, speech, sound, and music, achieving competitive performance on seen tasks and strong few-shot/zero-shot generalization on a wide range of unseen tasks.

\end{itemize}

\begin{figure*}[t]
    \centering
    \includegraphics[width=0.8\textwidth]{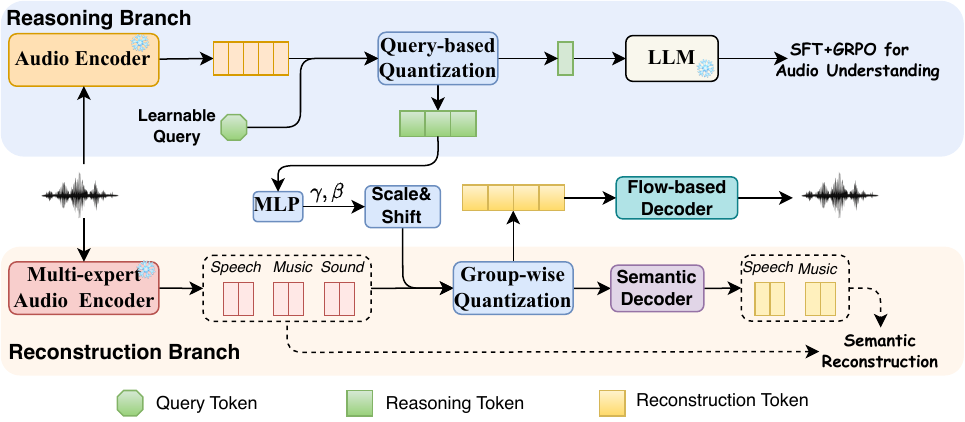}
    \caption{Figure X. Overview of the proposed ReasoningCodec framework. ReasoningCodec adopts a dual-branch architecture, consisting of a reasoning branch and a reconstruction branch, which are coupled through FiLM-based modulation. 
The semantic decoder consists of several convolutional layers.
}
    \label{fig:overview-codec}
\end{figure*}

\section{Related Works}
\subsection{Audio Language Models}
Recent years have witnessed rapid progress in audio language models that aim to bridge audio and text through multimodal learning \cite{qwen-omni,mimo-audio,borsos2023audiolm,kimi-audio}. Existing approaches largely fall into two distinct paradigms, depending on how audio representations are integrated with language models.
The first paradigm focuses on audio understanding by coupling continuous audio representations with pre-trained language models \cite{qwen2-audio,salmonn,qwen-omni}. In this line of work, audio signals are encoded into continuous features using pre-trained audio encoders \cite{whisper,wavlm,hubert}, which are then aligned with textual representations to support perception and reasoning tasks. 
The second paradigm formulates audio modeling as a discrete sequence prediction problem, drawing inspiration from autoregressive text language models \cite{borsos2023audiolm,valle,speartts,uniaudio}. In this setting, raw audio is first converted into a sequence of discrete tokens using an audio tokenizer or codec \cite{soundstream,encodec,dac,speechtokenizer,yang2023hifi}, and an autoregressive transformer is trained to model and generate audio token sequences. This paradigm has been successfully applied to text-to-speech~\citep{valle, speartts}, music generation \cite{musicgen,musiclm}, speech-to-speech dialogue \cite{moshi,glm4-voice,kimi-audio,spiritlm}, and unified multi-task audio generation models \cite{uniaudio,liu2025unitok,wang2024speechx,audiobox}. Recently, more work has focused on building unified speech understanding and generation models under the LM paradigm, such as OpusLM \cite{tian25b_interspeech}, DualSpeechLM \cite{dualspeechlm}, and Ming-UniAudio \cite{ming-uniaudio}. In this work, we focus on building unified audio understanding and generation models that can understand and generate text, speech, sound, and music. 

\subsection{Audio Tokenizer}
Audio tokenization is a key design choice in audio language models, as it determines the intermediate representation on which both understanding and generation models are built. Existing approaches can be broadly grouped into two families: \emph{continuous representations} and \emph{discrete tokens}, which offer fundamentally different trade-offs. \\
\textbf{Continuous representations}
A large body of work performs audio understanding by coupling continuous audio features with language models. In this paradigm, audio is encoded into continuous embeddings using HuBERT~\citep{hubert}, the Whisper encoder \cite{whisper}, WavLM~\citep{wavlm}, and so on. Continuous audio features preserve rich perceptual information and typically provide strong performance for understanding-oriented tasks. However, directly generating high-dimensional continuous sequences is less amenable to scalable autoregressive modeling: the large embedding dimension makes sequence prediction and sampling substantially more difficult compared to token-based generation, and often require additional modeling assumptions or specialized decoders \footnote{We note that many works focus on continuous autoregressive audio generation with diffusion decoder~\cite{rouard2025continuous}, where the modeling target is typically a low-dimensional VAE-based feature.}.  \\
\textbf{Discrete tokenizers}
To enable scalable autoregressive generation, many audio language models instead discretize audio into token sequences via clustering or vector quantization \cite{soundstream, encodec, dac}. Discrete representations are naturally compatible with language-model-style sequence prediction, but they vary significantly in the type of information retained. A first subclass constructs \emph{semantic discrete tokens} by quantizing intermediate representations of SSL encoders (e.g., via K-means or VQ)~\citep{glm4-voice, cosyvoice, semanticodec}. Prior work~\citep{borsos2023audiolm} suggests that semantic tokens are generally easier for language models to predict in generation tasks. 
While semantic discrete tokens capture partial high-level information, they often remain less effective than continuous semantic representations on language model based audio understanding tasks, due to the quantization-induced information bottleneck. A second subclass comprises \emph{acoustic discrete tokens} produced by neural audio codecs trained primarily for waveform reconstruction~\citep{soundstream, encodec, yang2023hifi, dac, siuzdak2023vocos,apcodec,simplespeech,taae,wu2024ts3,single-codec}. However, most acoustic tokenizers emphasize low-level fidelity and provide limited semantic abstraction, which can restrict their usefulness as intermediate representations for understanding. Furthermore, directly modeling the acoustic token with LM is also hard than modeling semantic token \cite{borsos2023audiolm}. \\
\textbf{Towards unified representations}
The above discussion highlights a persistent tension: continuous representations are strong for understanding but inconvenient for scalable generation, whereas discrete tokens enable generation but often impose an information bottleneck for understanding. 
Motivated by these limitations, we propose \textbf{ReasoningCodec}, an audio codec that explicitly factorizes audio into \emph{reasoning tokens} and \emph{reconstruction tokens}. This design aims to preserve language-aligned, reasoning-relevant information for understanding while maintaining sufficient acoustic structure for faithful reconstruction and high-quality autoregressive generation, thereby serving as a unified intermediate representation for both capabilities.

\section{ReasoningCodec}
\label{sec:reasoningcodec}
A core challenge for unified audio models is to design an intermediate audio representation that is simultaneously \emph{LM-friendly} for autoregressive modeling and \emph{information-preserving} for LM-based understanding. 
Prior work~\citep{borsos2023audiolm,almtokenizer} shows that \emph{semantic} discrete tokens are generally easier to model than purely acoustic tokens in autoregressive generation. 
However, directly using discrete semantic tokens for understanding remains non-trivial: vector quantization introduces information loss, which often degrades performance on comprehension tasks (see Table~\ref{tab:downstream_understanding}).
To bridge this gap, we propose \textbf{ReasoningCodec}, a factorized audio tokenizer that decomposes an audio waveform into two complementary token streams:
(i) \textbf{Reasoning tokens} \footnote{Note that reasoning tokens differ from conventional semantic tokens. While prior semantic tokenizers typically capture both text-level semantics and non-textual paralinguistic information such as prosody, our reasoning tokens are deliberately designed to capture only text-level semantic information. As a result, reasoning tokens contain less information than traditional semantic tokens and focus on high-level, text aligned semantics.}, which encode text-aligned, high-level perceptual analyses and planning representations, optimized to match the inductive biases of text LLMs for efficient understanding and generation.
Here, \emph{reasoning} denotes grounded perceptual inference over acoustic cues, rather than explicit multi-step chain-of-thought reasoning in text-based LLMs~\cite{guo2025deepseek}.
(ii) \textbf{Multi-level reconstruction tokens} that capture \emph{semantic content and fine-grained acoustics} for high-fidelity waveform reconstruction and LM-based autoregressive generation. 
Formally, given an audio $x$, ReasoningCodec produces $r = \mathcal{T}_r(x)$ and $s = \mathcal{T}_s(x \mid r)$, where $r$ denotes reasoning tokens and $s$ denotes multi-level reconstruction tokens. $\mathcal{T}_r$ and $\mathcal{T}_s$ denote the reasoning branch and reconstruction branch, respectively. Note that the waveform is reconstructed only from $\hat{x} = \mathcal{D}(s)$. Figure \ref{fig:overview-codec} shows the framework of ReasoningCodec. In the following, we present the details of the reasoning branch, the reconstruction branch, and the training procedure of ReasoningCodec. 
\subsection{Reasoning Branch}
\textbf{Encoders and architecture}
The reasoning branch uses multiple frozen pre-trained audio encoders to cover diverse domains: a Whisper encoder \cite{higg-audio} and a music SSL encoder \cite{muq}. Since the music encoder outputs $25$ Hz features, we downsample Whisper features to the same temporal rate before fusion. The query-based transformer encoder follows~\citep{almtokenizer} with $4$ transformer layers. Finally, a pre-trained text LLM (LLaMA-3.2-3B~\citep{llama3}) is used as the decoder head and updated with LoRA. \\
\textbf{Query-based compression and quantization.}
We obtain reasoning tokens via query-based quantization~\citep{almtokenizer}, which compresses audio into an extremely low frame-rate token sequence ($5$\,Hz). Let $h \in \mathbb{R}^{T\times d}$ be the continuous audio features extracted from pre-trained encoders. We use learnable queries $Q\in\mathbb{R}^{M\times d}$ and apply a lightweight transformer to summarize $h$ into query states:
\begin{equation}
\label{eq:query-attn}
z = \mathrm{Enc}\!\left(Q,\, h\right)\in \mathbb{R}^{M\times d},
\end{equation}
where $M = \lceil T/K\rceil$ is controlled by an interleaving factor $K$ (we set $K=5$). We then quantize $z$ using residual vector quantization (RVQ) with 8 codebooks. \\
\textbf{Training objectives}
Reasoning tokens are optimized to support fine-grained audio perception and text-aligned analytical abstractions. 
Specifically, we train the reasoning branch in two stages. 
First, we supervise the model on multiple audio understanding tasks (e.g., ASR and audio captioning) using supervised fine-tuning (SFT); detailed task configurations are provided in Appendix~\ref{appendix:codec_training_details}. 
Second, we further apply GRPO-style reinforcement learning~\citep{shao2024deepseekmath} to encourage detailed and grounded analytical descriptions of audio properties. 
Let $Y$ denote a sampled output and $R(Y)$ the corresponding reward computed by task-specific verifiers. 
In our implementation, we use LLaMA~3.1-Instruct~8B as the verifier to score the generated analyses based on their consistency with the final answers. 
Additional details are provided in Appendix~\ref{appendix:codec_training_details}.

\subsection{Reconstruction Branch} \label{sub_sec:semantic_branch}
\textbf{Motivation}
Reasoning tokens are aligned with text-level information and cannot be used to recover waveforms. We thus introduce a reconstruction branch that produces \emph{reconstruction tokens} while retaining structural semantics across speech, sound, and music. \\
\textbf{Multi-expert semantic features}
Following prior work on LM-friendly audio tokenizers~\citep{almtokenizer}, we use multiple frozen experts to extract semantically rich features that cover speech, music, and general sounds:
(i) WavLM for phone-level speech semantics;
(ii) a music SSL encoder for music-structure semantics;
(iii) a Whisper encoder for environmental sound semantics and residual acoustic cues.
Let these features be $h^{\text{ph}}, h^{\text{mu}}, h^{\text{env}}$. \\
\textbf{Group-wise quantization}
We quantize these features using three group-wise VQ modules. Specifically, we allocate one VQ layer for phone semantics, one for music structure, and $L_s^{\text{env}}=6$ layers for environmental sound semantics and remaining acoustic information:
\begin{equation}
\label{eq:groupvq}
s = \Big(
\mathrm{VQ}_1(h^{\text{ph}}),\
\mathrm{VQ}_2(h^{\text{mu}}),\
\mathrm{RVQ}_{3:8}(h^{\text{env}})
\Big).
\end{equation}
This yields a multi-level reconstruction token sequence $s$. \\
\textbf{Conditioning with reasoning tokens via FiLM}
To reduce redundancy between $r$ and $s$ and to inject reasoning-level context into the reconstruction process, following \cite{yang2025diffsoundstream}, we condition semantic features on reasoning tokens using feature-wise linear modulation (FiLM)~\citep{perez2018film}. Given semantic feature embeddings $S_e$ and quantized reasoning feature $\hat{R}$, we compute
\begin{equation}
\label{eq:film}
\mathrm{FiLM}(S_e; \hat{R}) = \gamma(\hat{R})\odot S_e + \beta(\hat{R}),
\end{equation}
where $\gamma(\cdot)$ and $\beta(\cdot)$ are small networks (implemented as lightweight MLPs). \\
\textbf{Decoder and training strategy}
We use a flow-based scalar latent diffusion decoder \citep{simplespeech} as the decoder (details in Appendix \ref{appendix:codec_training_stage3}). Specifically, we expect the decoder to predict the latent representations of SQ-Codec \cite{simplespeech} based on the reconstruction tokens. 
The reconstruction branch is trained with a flow-based objective and semantic feature reconstruction loss.
\begin{equation}
\label{eq:recon}
\mathcal{L}_{\mathrm{rec}}
= \mathcal{L}_{\mathrm{flow}}
+ \lambda_{\mathrm{sem}} \, \mathcal{L}_{\mathrm{sem}},
\end{equation}
\begin{equation}
\label{eq:flow}
\mathcal{L}_{\mathrm{flow}}
= \mathbb{E}_{t,\,\epsilon}\Big[
\big\|
v_\theta(z_t, t, s) - \epsilon
\big\|_2^2
\Big],
\end{equation}
where $\mathcal{L}_{\mathrm{sem}}$ denotes semantic feature matching terms adopted in prior work \cite{xcodec}. 
During training, we freeze the pre-trained experts and the reasoning branch, and update the VQ modules and the flow-based decoder.

\section{UniAudio 2.0}
\label{sec:uniaudio2}

Section~\ref{sec:reasoningcodec} introduces \textbf{ReasoningCodec}, which factorizes an audio waveform into (i) reasoning tokens that capture language-aligned abstractions for understanding and (ii) multi-level reconstruction tokens that preserve reconstruction-friendly information. Building on this tokenizer, we develop a unified multi-task audio foundation model, termed \textbf{UniAudio 2.0}. In the following, we describe the tokenization scheme, the unified vocabulary, the multi-stream input representation, the proposed functionally specialized autoregressive architecture, and multi-stage training strategy.
\subsection{Tokenization and Vocabulary}
UniAudio~2.0 supports two modalities: audio and text. 
For audio, we apply ReasoningCodec to obtain two token sequences: reasoning tokens $r$ and reconstruction tokens $s$. In our implementation, both $r$ and $s$ are represented with $K=8$ codebooks (i.e., $8$ parallel token streams per time step). 
For text, we adopt the tokenizer of the underlying pre-trained LLM and represent text as a single token stream. \\
\textbf{Joint vocabulary}
We build UniAudio~2.0 with a unified vocabulary that includes text tokens, reasoning tokens, and reconstruction tokens, together with special control symbols (e.g., \texttt{PAD}, \texttt{BOS}, \texttt{EOS}, and modality markers). 
Let the vocabulary sizes be $N_t$ (text), $N_r$ (reasoning), and $N_s$ (reconstruction). We initialize the text embedding from the pre-trained LLM, while audio-related embeddings are randomly initialized.

\subsection{Multi-Stream Representation}
\paragraph{Packing multi-modal sequences.}
To enable a single autoregressive transformer to process both modalities, we represent each time step as a \emph{multi-stream token vector}. 
Let $K=8$ denote the number of audio codebooks and let the last stream index $K$ be reserved for text. 
We form a $S=9$ stream representation, where the first $K$ streams are audio and the last stream is text. Concretely, we construct an input token tensor $X \in \mathbb{Z}^{B \times T \times S}$ where $B$ is the batch size, $T$ is the packed sequence length, and $S=9$ is the number of streams. For a text position, we place a text token in the last stream and set all audio streams to \texttt{PAD}. 
For an audio position, we place audio tokens in the first $K$ streams and set the text stream to \texttt{PAD}.
This design allows a single transformer to consume heterogeneous sequences without changing the backbone architecture. \\
\textbf{Stream-wise embeddings and fusion}
We assign a separate embedding table to each stream and fuse them by masked summation. 
Let $x_{t,i}$ denote the token at time step $t$ and stream $i$.
We define a binary mask $m_{t,i}\in\{0,1\}$ indicating whether $x_{t,i}$ is a valid (non-\texttt{PAD}) token. 
The fused token representation is computed as $h_t = \sum_{i=1}^{S} m_{t,i}\, E_i(x_{t,i})$
where $E_i(\cdot)$ is the embedding lookup for the $i$-th stream.
In practice, only one modality is active at each time step.

\subsection{Unified Autoregressive Architecture}
\textbf{Backbone initialization}
UniAudio~2.0 is initialized from a pre-trained text LLM (LLaMA~3.2~3B~\citep{llama3}) to inherit strong text knowledge.
To incorporate audio perception and audio generation capabilities while retaining a unified autoregressive interface, we introduce a functionally specialized architecture. \\
\textbf{Layer specialization.}
Let the whole transformer backbone consist of three consecutive blocks:
\begin{equation}
\label{eq:fuse-ar}
H^{(u)} = \mathcal{F}_{\mathrm{u}}(h),\quad
H^{(c)} = \mathcal{F}_{\mathrm{cm}}(H^{(u)}),\quad
H^{(g)} = \mathcal{F}_{\mathrm{g}}(H^{(c)})
\end{equation}
where $\mathcal{F}_{\mathrm{u}}$ denotes \textbf{audio understanding experts} (lower layers), 
$\mathcal{F}_{\mathrm{cm}}$ denotes \textbf{cross-modal experts} (middle layers),
and $\mathcal{F}_{\mathrm{g}}$ denotes \textbf{audio generation experts} (upper layers).
The cross-modal experts are initialized from the pre-trained LLM to preserve textual knowledge, while the audio-specific experts are randomly initialized. \\
\textbf{Audio-only computation in specialized experts}
A key design is that both audio understanding experts and audio generation experts operate \emph{exclusively} on audio streams, leaving text tokens unchanged. 
Let $M_{\mathrm{aud}} \in \{0,1\}^{B\times T}$ be a binary mask indicating whether a position corresponds to audio (i.e., the text stream is \texttt{PAD}). 
For a transformer block output $f(\cdot)$, we implement audio-only updates as
\begin{equation}
\label{eq:audio-only}
H' = H + M_{\mathrm{aud}}\odot \big(f(H) - H\big)
\end{equation}
which updates hidden states only at audio positions and keeps text positions intact. This mechanism preserves the pre-trained text processing pathway while enabling dedicated capacity for audio perception and synthesis. \\
\textbf{Autoregressive modeling.}
UniAudio~2.0 is trained under a unified autoregressive framework over the packed multi-stream sequence. 
While text and audio tokens are modeled within a single transformer backbone, their prediction heads are different. The text tokens are predicted at the token level following standard language modeling practice. 
The text autoregressive loss is defined as
\begin{equation}
\label{eq:text-loss}
\mathcal{L}_{\mathrm{text}}
= - \sum_{t \in \mathcal{T}}
\log p_\theta(x_{t,\mathrm{text}} \mid X_{<t}).
\end{equation}
where $X$ denotes the multi-stream sequence. $\mathcal{T}$ denotes the set of text token positions.
Audio tokens are modeled at the frame level. Each audio frame corresponds to $K$ parallel reconstruction or reasoning tokens. Rather than predicting these tokens directly from the backbone, we follow the local autoregressive decoding strategy introduced in \cite{uniaudio,moshi}, and employ a lightweight audio decoder conditioned on the hidden states from the audio generation experts, $H^{(g)}=\{h^{(g)}_t\}_{t=1}^T$.
For an audio frame at time step $t$, the local decoder autoregressively predicts the $K$ audio tokens.
The audio autoregressive loss is then given by
\begin{equation}
\label{eq:audio-loss}
\mathcal{L}_{\mathrm{audio}}
= - \sum_{t \in \mathcal{A}}
\sum_{k=1}^{K}
\log p_\theta(x_{t,k} \mid x_{t,<k},\, h^{(g)}_t),
\end{equation}
with $\mathcal{A}$ denoting the set of audio frame positions. \\

\textbf{Overall training objective}
The final autoregressive objective for UniAudio~2.0 combines text and audio losses:
\begin{equation}
\label{eq:overall-ar}
\mathcal{L}_{\mathrm{AR}}
= \lambda_{\mathrm{text}} \mathcal{L}_{\mathrm{text}}
+ \lambda_{\mathrm{audio}}\, \mathcal{L}_{\mathrm{audio}},
\end{equation}
where $ \lambda_{\mathrm{text}}$ and $\lambda_{\mathrm{audio}}$ balances the contributions of the two modalities. 
Details of the multi-stage training procedure are provided in Section~\ref{sec:training}.
\subsection{Data and Training Strategy}
\label{sec:training}

UniAudio~2.0 is trained on a diverse collection of text and audio data under a multi-task, multi-stage training paradigm. Our data construction and training strategy are designed to: (i) preserve the strong textual capability inherited from pre-trained LLMs, (ii) progressively inject audio understanding and generation abilities, and (iii) improve generalization to unseen tasks.

\textbf{Multi-task Data Construction.}
We organize the training corpus into several complementary data types, each corresponding to a class of tasks supported by UniAudio~2.0. The training data include: (1) text-only data, (2) audio-only data, (3) speech--transcription paired data, (4) speech--caption--transcription paired data, (5) audio/music--caption paired data, (6) lyric--song paired data, and (7) auditory sentences constructed using our proposed task-construction strategy. Detailed descriptions of the data and tasks are provided in Appendix~\ref{appendix:uniaudio_training_data}.

\textbf{Multi-stage Training Strategy.}
We adopt a four-stage training strategy to progressively integrate audio understanding and generation capabilities into the unified autoregressive model, including: (1) Stage~1: Audio understanding warm-up, (2) Stage~2: Audio generation warm-up, (3) Stage~3: Audio--text pre-training, and (4) Stage~4: Audio--text mid-training. Figure~\ref{fig:overview-training-stage} provides an overview of the four training stages.

\textbf{Stage~1: Audio Understanding Warm-up.}
In the first stage, we focus on initializing the audio understanding experts. The model is trained on a subset of audio understanding tasks while all other components are frozen. To encourage the understanding experts to encode rich semantic information, we introduce an auxiliary semantic distillation objective. Following the training of ReasoningCodec (Section~\ref{sub_sec:semantic_branch}), a lightweight decoder is attached to reconstruct semantic features extracted from frozen WavLM and music SSL models. The overall objective consists of a reconstruction loss and a language modeling loss. After training, the auxiliary decoder is discarded.

\textbf{Stage~2: Audio Generation Warm-up.}
In this stage, we train the audio generation expert and the local audio decoder. The model is optimized on a subset of audio generation tasks, while the understanding and cross-modal experts remain fixed.

\textbf{Stage~3: Audio--Text Pre-training.}
We jointly update all model parameters using a mixture of audio understanding tasks, audio generation tasks, text-only data, and audio-only data. This stage aligns the two modalities under a unified autoregressive objective. The maximum context length in this stage is 1024.

\textbf{Stage~4: Audio--Text Mid-training.}
In the final stage, we aim to extend the effective context length and enhance generalization to unseen tasks. We continue training on a subset of the pre-training data from Stage~3, augmented with the constructed auditory sentence data. This stage encourages the model to reason over longer and more complex audio--text sequences and improves robustness across diverse task settings. The maximum context length in this stage is 2048.

Additional details of each training stage are provided in Appendix~\ref{appendix:uniaudio2_4_stage_training}.

\begin{table*}[t]
\centering
\setlength{\tabcolsep}{2pt}
\caption{Unified codec reconstruction results on speech, sound, and music. PESQ is reported in two variants: wideband (WB) and narrowband (NB). The AudioBox score includes CE, CU, PC, and PQ. Since some tokenizers (e.g., DAC and X-Codec) operate at a higher frame rate, we report comparisons under the same token-rate setting by adjusting the number of RVQ layers for the baselines.
}
\begin{tabular}{cccccccccc}
\hline
\multirow{2}{*}{\textbf{Model}} &
\multicolumn{5}{c}{\textbf{Speech}} &
\multicolumn{2}{c}{\textbf{Sound}} &
\multicolumn{2}{c}{\textbf{Music}} \\
\cmidrule(lr){2-6}\cmidrule(lr){7-8}\cmidrule(lr){9-10}
& \textit{PESQ} & \textit{STOI} & \textit{UT-MOS} & \textit{VISQOL} & \textit{SIM}
& \textit{VISQOL} & \textit{AudioBox Score}
& \textit{VISQOL} & \textit{AudioBox Score} \\
\hline
DAC-Codec     & 2.10/2.29 & 0.81 & 3.13 & 3.67 & 0.91 & 3.02 & 3.34/4.25/3.78/5.44 & \textbf{4.06} & 6.98/6.97/6.25/7.09 \\
EnCodec       & 2.00/2.24 & 0.81 & 2.58 & 3.64 & 0.92 & 2.99 & 3.61/4.62/3.82/5.49 & 4.04 & 6.60/6.57/\textbf{6.27}/6.71 \\
MimiCodec     & 2.09/2.82 & 0.82 & 3.65 & 3.82 & 0.96 & -- & -- & -- & -- \\
Higg-Audio    & 2.20/2.90 & 0.78 & 3.90 & 3.84 & 0.96 & \textbf{3.20} & 4.01/4.97/3.65/5.87 & 4.01 & 6.95/7.48/5.01/7.65 \\
X-Codec       & 2.08/2.72 & 0.83 & 3.75 & 3.90 & 0.94 & 3.01 & 3.97/4.82/\textbf{3.98}/5.87 & 3.82 & 7.43/7.19/{6.21}/7.24 \\
ALMTokenizer  & 2.00/2.30 & 0.81 & 3.76 & 3.78 & 0.92 & 2.99 & 4.02/4.65/3.24/5.66 & 3.96 & 6.44/6.68/6.12/6.94 \\
Ours          & \textbf{2.36}/\textbf{2.93} & \textbf{0.85} & \textbf{3.91} & \textbf{3.94} & \textbf{0.97} & 3.10 & \textbf{4.12}/\textbf{5.06}/3.58/\textbf{5.96} & 4.03 & \textbf{7.51}/\textbf{7.68}/6.12/\textbf{7.87} \\
\hline
\end{tabular}
\label{tab:codec_recon_all}
\end{table*}

\begin{table*}[t]
\centering
\setlength{\tabcolsep}{5pt}
\caption{LLM-based perplexity (PPL) across codebooks. We present the results from first four VQ layers. Following the common practice, lower PPL denotes better performance. Note that all of tokenizers have the same codebook size (1024).}
\begin{tabular}{ccccccccccc}
\hline
\multirow{2}{*}{Model} &
\multicolumn{5}{c}{Speech} &
\multicolumn{5}{c}{Music} \\
\cmidrule(lr){2-6}\cmidrule(lr){7-11}
& VQ1 & VQ2 & VQ3 & VQ4 & Avg
& VQ1 & VQ2 & VQ3 & VQ4 & Avg \\
\hline
XCodec \cite{xcodec}            & 9.22 & 16.31& 21.73 & 26.36 & {18.40} & 11.77& 23.60& 34.47 & 42.3  & {28.5} \\
HiggCodec \cite{higg-audio}        & 9.12 & 29.49& 51.32 & 65.92 & {38.46} & 12.83& 21.34& 57.82& 88.1  & {45.02} \\
DAC Codec \cite{dac}        & 13.74& 66.06& 116.54& 163.89& {90.06}  & 25.67& 68.09& 99.66 & 133.9 & {81.34} \\
\midrule
Reason-only (Ours)           & \textbf{3.68} & \textbf{7.73} & \textbf{15.88} & \textbf{18.99} & \textbf{11.57} & \textbf{3.81} & \textbf{6.97} & \textbf{7.26}  & \textbf{8.20}  & \textbf{6.56} \\
Reconstruction-only (Ours)          & 8.69 & 13.99& 19.75 & 75.02 & {29.36} & 9.19 & 22.94& 15.94 & 59.3  & {26.33} \\
Reason + Reconstruction (Ours) & 6.92 & 11.37& 19.42 & 55.36 & {23.77} & 7.02 & 15.55& 15.23 & 42.5  & {20.1} \\
\hline
\end{tabular}
\label{tab:ablation_llm_ppl_speech_music}
\end{table*}

\begin{table}[t]
\centering
\caption{Downstream understanding evaluation on LLM-based ASR (ASR), emotion classification (ER), audio classification (AC), and music classification (MC).}
\begin{tabular}{ccccc}
\hline
Model & ASR & ER & AC & MC \\
\hline
Whisper & \textbf{8.5} & \textbf{59.2} & 54.3 & \textbf{74} \\
DAC & 93.2 & 5.2 & 19 & 28 \\
XCodec & 37.6 & 29.4 & 44.7 & 46 \\
Higg-Codec & 31.2 & 30 & 49.4 & 38 \\
ALMTokenizer & 26.9 & 32.4 & 50.1 & 45 \\
Reason-only & 10.1 & 50.2 & 33 & 80 \\
Reconstruction-only & 16.3 & 42.1 & 48.7 & 65 \\
Reason+Reconstruction & 9.0 & 56.4 & \textbf{63.3} & 70 \\
\hline
\end{tabular}
\label{tab:downstream_understanding}
\end{table}

\begin{table*}[t]
    \centering
    \small
    \caption{The subjective reconstruction results using MUSHRA (comparative scoring of samples) of codec models on speech, sound and music. \textbf{Bold} for the best result. FPS denotes that the frame number in one second. TPS denotes that the token number in one second.}
    \vspace{2mm}
    \scalebox{0.99}{
    \begin{tabular}{lcccccc}
    \toprule
    Models & FPS/TPS &Cookbook size  & Speech ($\uparrow$)   & Sound ($\uparrow$) & Music ($\uparrow$)   \\
    \midrule
    MimiCodec (8 RVQ) \cite{moshi}  & 12.5/100 & 2048 & 86.7 $\pm$ 2.1 & -   & -      \\ 
    XCodec \cite{xcodec}  & 50/100 & 1024 & 78.5 $\pm$ 4.5 &  72.6 $\pm$ 2.1  & 69.8 $\pm$ 1.9      \\
    Higgs-Audio \cite{higg-audio}  & 25/100 & 1024 & 84.4 $\pm$ 2.6 &  79.2 $\pm$ 1.8   & 81.0 $\pm$ 1.6      \\
    Encodec  \cite{encodec}   & 75/150 & 1024 & 69.3 $\pm$ 2.4 &  68.5 $\pm$ 2.0  &  62.6 $\pm$ 2.2   \\
    DAC \cite{dac}   & 50/100 & 1024 & 71.3 $\pm$ 1.9 &  {70.0 $\pm$ 1.9}  & 63.0 $\pm$ 1.8     \\
    \midrule
    ReasoningCodec (Ours) & 12.5/100 & 1024 & \textbf{90.5 $\pm$ 2.8} & \textbf{80.8 $\pm$ 2.0}  & \textbf{86.6 $\pm$ 2.3} \\
    \bottomrule 
    \end{tabular}
    \label{tab:mos}}
\end{table*}

\subsection{Connection to the UniAudio Series}

UniAudio~2.0 continues the UniAudio research line~\cite{uniaudio,uniaudio1.5}, which aims to build a unified foundation model for diverse audio understanding and generation tasks.
Compared with previous versions, UniAudio~2.0 introduces substantial advances in representation learning, model architecture, and training paradigm.

\paragraph{Goal.}
Similar to UniAudio and UniAudio~1.5, UniAudio~2.0 aims to develop a general-purpose audio foundation model that supports speech, sound, and music understanding and generation within a unified framework.
While earlier systems primarily focused on multi-task learning and in-context adaptation, UniAudio~2.0 further emphasizes representation-level alignment between audio and text, enabling more scalable and transferable modeling.
\paragraph{Representation.}
UniAudio and UniAudio~1.5 mainly relied on acoustic codecs or LLM-driven tokenization schemes.
In contrast, UniAudio~2.0 introduces ReasoningCodec, a factorized audio tokenizer that explicitly separates text-aligned reasoning tokens from reconstruction tokens.
This design provides stronger semantic abstraction for understanding while preserving fine-grained acoustic information for high-fidelity generation.
\paragraph{Architecture.}
Previous versions adopted decoder-only transformer backbones without considering the cross-modal fusion.
UniAudio~2.0 employs a unified autoregressive backbone with specialized understanding, generation, and cross-modal experts, allowing more stable and scalable joint optimization across heterogeneous tasks.
\paragraph{Training paradigm.}
Earlier systems were mainly trained using supervised multi-task learning and limited-scale pre-training.
UniAudio~2.0 adopts a multi-stage training pipeline and large-scale audio-text pre-training. This paradigm substantially improves few-shot and zero-shot generalization.
\paragraph{Capabilities.}
Benefiting from these advances, UniAudio~2.0 extends previous systems from primarily in-domain performance to stronger cross-task and cross-domain generalization.
It demonstrates improved robustness on complex understanding tasks, more controllable generation, and enhanced adaptability to unseen scenarios.

Overall, UniAudio~2.0 preserves the unification philosophy of the UniAudio series while advancing its core representation and learning paradigm to a new level of scalability and generalization.

\section{Experiments}

\begin{table*}[t]
\centering
\small
\setlength{\tabcolsep}{8pt}
\renewcommand{\arraystretch}{1.25}
\caption{The performance comparison on Seen tasks between previous SOTA models and UniAudio 2.0. For each task, we choose the commonly used benchmark and metrics. Note that we list the most representative and related works with us, the more comprehensive comparison can be found in Appendix \ref{appendix:uniaudio_detailed_exp}. \textbf{Bold} for the best result.}
\begin{tabular}{p{0.36\textwidth} p{0.28\textwidth} p{0.30\textwidth}}
\toprule
\multicolumn{1}{c}{\textbf{Task \& Datasets}} &
\multicolumn{1}{c}{\textbf{Model}} &
\multicolumn{1}{c}{\textbf{Performance}} \\
\midrule

\makecell[c]{\textbf{TTS} \\ ZH / EN / LS-clean} &
\makecell[c]{MiMo-Audio-7B-Instruct \\ Qwen2.5-Omni 7B \\ UniAudio 2.0 (Ours) } &
\makecell[c]{ 1.93 / 5.37 / 4.74 \\
\textbf{1.21} / \textbf{3.10} / 4.28 \\
2.30 / 3.63 / \textbf{3.46} } \\
\midrule

\makecell[c]{\textbf{InstructTTS} \\ WER/ Style-Acc \\
    / UTMOSv2} &
\makecell[c]{MiMo-Audio-7B-Instruct \\ CapSpeech-AR \\ UniAudio 2.0 (Ours)} &
\makecell[c]{ 7.8 / 40.5 / 3.17 \\
              9.1 /	\textbf{52.2} / 3.18 \\
              \textbf{7.3} /	42.3 /	\textbf{3.38} } \\
\midrule

\makecell[c]{\textbf{ASR}\\ LS-clean / LS-other \\
/ Seed-EN / Seed-ZH} &
\makecell[c]{MiMo-Audio-7B-Instruct\\Qwen2.5-Omni-7B \\ UniAudio 2.0 (Ours)} &
\makecell[c]{ 3.5 /	35.4 /	29.8 /	7.0 \\
              3.9 / \textbf{5.5} /	\textbf{1.3} /	2.9 \\ 
              \textbf{2.7} /	6.3 / 	2.6 / 	\textbf{2.1} } \\
\midrule

\makecell[c]{\textbf{Audio Caption \& Generation}\\ CIDER / KL / FD} &
\makecell[c]{Qwen2.5-Omni \\ Stable Audio Open  \\ UniAudio 2.0} &
\makecell[c]{
             0.39 / - / -  \\
             - / \textbf{2.14} / 78.2  \\
             \textbf{0.69} / 3.26 / \textbf{50.7} } \\
\midrule

\makecell[c]{\textbf{Music Caption \& Generation}\\ GPT-score / KL / FAD} &
\makecell[c]{Qwen2.5-Omni-7B \\ MusicGen  \\ UniAudio 2.0} &
\makecell[c]{
             \textbf{5.33} / - / -  \\
             - / \textbf{1.31} / 5.0 \\ 
             5.14 / 1.8 / \textbf{3.44} } \\
\midrule

\makecell[c]{\textbf{Song Generation}\\ WER / CE/CU/PQ} &
\makecell[c]{SongGen \\ UniAudio 2.0} &
\makecell[c]{ 40.58 / 6.77 / 6.86 / 7.19\\
              \textbf{36.5} / \textbf{6.87} / \textbf{7.41} / \textbf{7.62} } \\
\midrule
\makecell[c]{\textbf{Lyric Recognition}\\ WER} &
\makecell[c]{Qwen2.5-Omni-7B \\UniAudio 2.0} &
\makecell[c]{ 56.99 \\
             \textbf{28.57} } \\

\bottomrule
\end{tabular}
\label{tab:seen_tasks}
\end{table*}

\begin{table*}[t]
\centering
\caption{Few-shot results across tasks. MOS denotes the DNS-MOS score. ACC denotes the accuracy.}
\label{tab:fewshot_all}
\setlength{\tabcolsep}{6pt}
\renewcommand{\arraystretch}{1.12}
\begin{tabular}{ccccc}
\toprule
\textbf{Task} & \textbf{Metrics} & \textbf{Model} & \textbf{1-shot} & \textbf{2-shot} \\
\midrule

\multirow{2}{*}{SE} &
\multirow{2}{*}{PESQ/STOI/WER/MOS} &
MiMo-Audio & 1.20 / 0.21 / 65.29 / 3.46 & 1.16 / 0.21 / 29.29 / 3.48 \\
& & Ours      & \textbf{1.55} / \textbf{0.64} / \textbf{14.13} / \textbf{3.76} & \textbf{1.52} / \textbf{0.66} / \textbf{14.82} / \textbf{3.77} \\
\midrule

\multirow{2}{*}{VC} &
\multirow{2}{*}{WER/SIM/MOS} &
MiMo-Audio & 20.95 / \textbf{0.90} / \textbf{3.80} & \textbf{14.05} / \textbf{0.93} / \textbf{3.78} \\
& & Ours      & \textbf{18.61} / 0.89 / 3.74 & 19.01 / 0.90 / 3.71 \\
\midrule

\multirow{3}{*}{Emotion} &
\multirow{3}{*}{EN / ZH ACC (\%)} &
UniAudio 1.5 & 45.0 / 46.6 & 52.0 / 51.0 \\
& & MiMo-Audio   & 42.5 / 45.0 & 53.6 / 52.4 \\
& & Ours         & \textbf{67.0} / \textbf{59.8} & \textbf{70.2} / \textbf{62.8} \\
\midrule

\multirow{3}{*}{Sound} &
\multirow{3}{*}{ACC (\%)} &
UniAudio 1.5 & 48.0 & 55.2 \\
& & MiMo-Audio   & 45.3 & 76.0 \\
& & Ours         & \textbf{59.8} & \textbf{62.8} \\
\bottomrule
\end{tabular}
\end{table*}

\begin{table}[t]
\centering
\caption{Zero-shot results across tasks. Metrics: MMLU reports Acc (\%); S2S denotes speech-to-speech/text instruction-following, we report S2S/S2T GPT-score; DSR reports WER (\%); A-I-TTS reports SIM / Style-Acc (\%) / WER (\%) / UTMOSv2; Speech$\rightarrow$Sound reports WER (\%) / CLAP-score / UTMOSv2. A-I-TTS denotes audio and caption guided speech generation. Speech-S denotes generate speech and sound, we use WER/CLAP-score/UTMOSv2 as the metrics.}
\label{tab:zeroshot_all}
\setlength{\tabcolsep}{6pt}
\renewcommand{\arraystretch}{1.12}
\begin{tabular}{c c c}
\toprule
\textbf{Task} & \textbf{Model} & \textbf{Score} \\
\midrule

\multirow{3}{*}{Text}
& LLaMA 3.2 1B & 34.14 \\
& LLaMA 3.2 3B & \textbf{47.63} \\
& Ours          & 44.10 \\
\midrule

\multirow{3}{*}{S2S}
& LLAMA-Omni & \textbf{3.47} / \textbf{3.99} \\
& SpeechGPT  & 2.19 / 2.98 \\
& Ours       & 2.16 / 3.66 \\
\midrule

\multirow{2}{*}{DSR}
& Qwen2.5-Omni & 80.6 \\
& Ours         & \textbf{19.4} \\
\midrule

A-I-TTS
& Ours & 0.89 / 32.62 / 11.57 / 2.87 \\
\midrule

Speech-S
& Ours & 6.15 / 0.11 / 2.96 \\
\bottomrule
\end{tabular}
\end{table}

\subsection{Dataset}
\textbf{Data preparation for the ReasoningCodec} ReasoningCodec is trained on approximately 10,000 hours of data. In the speech domain, we utilize a subset of Multilingual LibriSpeech (MLS) \cite{pratap2020mls}, with 5,000 hours randomly selected. In the sound domain, we utilize a subset of AudioSet, with 3,000 hours randomly selected; in the music domain, we employ a subset of the Million Song Dataset \cite{Bertin-Mahieux2011}, also with 2,000 hours randomly selected. Following \cite{almtokenizer}, we evaluate the codec’s speech reconstruction performance using a subset of the VCTK dataset \cite{vctk}, and assess both audio and music reconstruction performance using the AudioCaps \cite{audiocaps} validation set and the MusicCaps dataset \cite{musiclm}, respectively.

\textbf{Data preparation for audio language models} As we discussed in Section \ref{sec:uniaudio2}, the training data of UniAudio 2.0 includes multiple source. We list the data sources and detailed statistics in Appendix \ref{appendix:uniaudio_training_data}.  
\subsection{Evaluation Metrics}

\textbf{Audio Tokenizer Evaluation.}
We evaluate audio tokenizers from three perspectives: (1) audio reconstruction, (2) LLM-based audio understanding, and (3) LLM-based audio generation.

\textit{Audio Reconstruction.}
For speech reconstruction, we adopt DNS-MOS, UT-MOS, PESQ, STOI (Short-Time Objective Intelligibility), and VISQOL. For sound and music evaluation, we follow~\cite{higg-audio} and use VISQOL (audio version) and the AudioBox aesthetics score. In addition, following~\cite{dac}, we conduct MUSHRA subjective evaluations on speech, sound, and music. As shown in Table~\ref{tab:mos}, ReasoningCodec achieves consistently strong reconstruction performance across all modalities.

\textit{LLM-based Audio Understanding Tasks.}
To assess whether discrete audio tokenizers are suitable as intermediate representations for LLM-based audio understanding, we follow the settings of Qwen-Audio~\cite{qwen2-audio,salmonn} and conduct multi-task training on ASR~\cite{librispeech}, emotion recognition~\cite{esd}, audio classification~\cite{tut2017}, and music classification~\cite{music_genre}. During downstream task training, only the adapter and LoRA modules~\cite{hu2021lora} are updated.

\textit{LLM-based Audio Generation Tasks.}
To evaluate whether the proposed tokenizer is suitable for autoregressive modeling, we follow~\cite{almtokenizer,audiocodecbench} and adopt perplexity (PPL) and token prediction accuracy as evaluation metrics.

\textbf{Audio Understanding and Generation Tasks Evaluation.}
We evaluate UniAudio~2.0 on a diverse set of audio-related understanding and generation tasks. To comprehensively assess its capabilities, we consider three evaluation settings: (1) seen-task evaluation, (2) few-shot evaluation, and (3) zero-shot evaluation. Detailed task descriptions are provided in Appendix~\ref{appendix:uniaudio2_eval_data}.

\textbf{Seen Tasks.}
These tasks are observed during pre-training, including TTS, instructed TTS, ASR, audio generation and captioning, music generation and captioning, song generation, and lyric recognition. For these tasks, we follow commonly adopted benchmarks and evaluation metrics.

\textbf{Few-shot Tasks.}
We design several few-shot tasks, including speech denoising, voice conversion, emotion recognition, and audio classification. For each task, we consider both one-shot and two-shot settings. We note that MiMo-Audio~\cite{mimo-audio} also introduces a few-shot evaluation protocol; however, due to our context length limitation of 2048, we are unable to adopt the same experimental setup.

\textbf{Zero-shot Tasks.}
Finally, we evaluate UniAudio~2.0 on a set of unseen tasks to assess its zero-shot generalization ability, including text question answering, speech-to-speech conversation, dysarthric speech recognition, speech-to-sound generation, and audio-prompted instruction-following TTS. For these tasks, we do not include any corresponding task-specific training data, and therefore treat them as zero-shot evaluations.

\subsection{Performance of the Audio Tokenizer}

\textbf{Reconstruction Performance.}
Table~\ref{tab:codec_recon_all} presents the reconstruction performance on speech, sound, and music evaluation sets. We compare our method with previous state-of-the-art universal audio codecs, including DAC~\cite{dac}, Encodec~\cite{encodec}, Higg-AudioCodec~\cite{higg-audio}, X-Codec~\cite{xcodec}, and ALMTokenizer~\cite{almtokenizer}. We observe that the proposed ReasoningCodec achieves strong reconstruction quality across different audio modalities. In Table~\ref{tab:mos},, we further conduct subjective evaluations of different audio tokenizers. We can see that at the same token rate (TPS), ReasoningCodec achieves consistently better reconstruction performance.

\textbf{Token Modeling Performance.}
We investigate whether ReasoningCodec is suitable for LLM-based audio generation and understanding tasks. For generation experiments, we use the same LLM backbone, training data, and evaluation data, while varying only the audio tokenizer. As shown in Table~\ref{tab:ablation_llm_ppl_speech_music}, we observe that: (1) semantically enhanced audio tokenizers (e.g., X-Codec \cite{xcodec}) perform better than purely acoustic tokenizers (e.g., DAC-Codec \cite{dac}); (2) our proposed ReasoningCodec achieves strong performance in terms of perplexity and token prediction accuracy; and (3) reasoning tokens are easier for language models to capture, and combining reasoning tokens with reconstruction tokens significantly improves the prediction accuracy of reconstruction tokens. These results further demonstrate the effectiveness of ReasoningCodec.

Table~\ref{tab:downstream_understanding} reports the results on audio understanding tasks. The results show that ReasoningCodec achieves the best understanding performance among discrete audio tokenizers, and its performance is close to that of a continuous tokenizer (Whisper) on multiple benchmarks. Figure~\ref{fig:codec_understanding_loss} illustrates the training loss for different tokenizers. Notably, the training loss of reasoning tokens decreases rapidly, and these phenomena further validate the effectiveness of ReasoningCodec. 
In Section~\ref{sec:exp_seen_task}, we further show that UniAudio~2.0 built on ReasoningCodec achieves competitive performance compared with other audio understanding systems. In Appendix~\ref{appendix:codec_why_reason}, we provide a more detailed analysis of why introducing reasoning tokens is effective.

\subsection{The Performance on Seen Tasks} \label{sec:exp_seen_task}
In this section, we evaluate UniAudio~2.0 on tasks seen during pre-training, spanning speech, general audio, and music/song domains. We follow standard benchmarks and evaluation protocols, and compare against both unified audio–speech foundation models (e.g., MiMo-Audio~\cite{mimo-audio} and Qwen2.5-Omni-7B~\cite{qwen-omni}) and strong task-specific systems (e.g., MusicGen~\cite{musicgen}), where applicable.\footnote{Although many tasks have strong task-specific baselines, we adopt unified base models to enable fair and consistent comparisons across tasks.}

As summarized in Table~\ref{tab:seen_tasks}, we observe that: (1) UniAudio~2.0 achieves strong performance on both speech generation and recognition tasks (e.g., ASR, TTS, and instructed TTS), and supports multiple languages, including English, Mandarin Chinese, and Cantonese (see Appendix~\ref{appendix:uniaudio_detailed_exp}). Compared with the previous state-of-the-art speech foundation model, MiMo-Audio~7B, UniAudio~2.0 achieves better performance on both ASR and TTS tasks with only 3B parameters; (2) for general audio, UniAudio~2.0 remains competitive with previous state-of-the-art task-specific models (e.g., Stable Audio), and also outperforms Qwen2.5-Omni~7B on audio captioning tasks; and (3) for the music modality, it effectively supports both music/song understanding and generation tasks. Overall, UniAudio~2.0 consistently performs well across multiple benchmarks.

\subsection{Performance on Few-shot Tasks}

In this section, we investigate the generalization ability of UniAudio~2.0 to unseen tasks. Following the setting of GPT-3~\cite{gpt-3}, we evaluate our model on a series of few-shot tasks covering both generation and understanding.

For few-shot generation, we follow~\cite{uniaudio1.5,mimo-audio} and evaluate speech denoising and voice conversion. For few-shot understanding, we consider sound classification and emotion classification. Table~\ref{tab:fewshot_all} summarizes the experimental results, showing that UniAudio~2.0 achieves strong generalization performance on most benchmarks, especially in the 1-shot setting.


\begin{table*}[t]
\centering
\caption{Ablation studies on multi-stage training, layer specialization, and model size. Due to space limitations, we present only representative tasks in this table. More results are provided in the Appendix Table \ref{tab:appendix_ablation_more}.}
\label{tab:ablation_stage}
\setlength{\tabcolsep}{6pt}
\renewcommand{\arraystretch}{1.15}
\begin{tabular}{c c c c c c c}
\toprule
\textbf{Setting} & \textbf{MMLU} & \textbf{ASR} & \textbf{TTS} & \textbf{Audio Gen} & \textbf{Music Gen} & \textbf{Song Gen} \\
\midrule
w/o Stage 4
& 38.77
& \makecell[c]{LS-clean: 3.76 \\ LS-other: 7.38 \\ SEED-ZH: 4.3 \\  SEED-EN: 3.87}
& \makecell[c]{SEED-EN: 6.21 \\ SEED-ZH: 1.97}
& \makecell[c]{KL: 3.08 \\ FD: 47.6}
& \makecell[c]{KL: 1.87 \\ FAD: 3.66}
& \makecell[c]{WER: 37.4 \\
               CE/PQ: 6.8/7.6}
\\
\midrule
w/o Experts
& 37.2
& \makecell[c]{LS-clean: 4.17 \\ LS-other: 8.0 \\ SEED-ZH: 5.1 \\ SEED-EN: 4.22}
& \makecell[c]{SEED-EN: 5.31 \\ SEED-ZH: 2.76}
& \makecell[c]{KL: 3.98 \\ FD: 60.8}
& \makecell[c]{KL: 4.12 \\ FAD: 6.98}
& \makecell[c]{WER: 42.5 \\
               CE/PQ: 6.68 / 7.55}
\\
\midrule
1B
& 30.2
& \makecell[c]{LS-clean: 3.8 \\ LS-other: 7.4 \\ SEED-ZH: 3.5 \\ SEED-EN: 3.3}
& \makecell[c]{SEED-EN: 4.2 \\ SEED-ZH: 5.3}
& \makecell[c]{KL: 3.78 \\ FD: 82.4}
& \makecell[c]{KL: 4.3 \\ FAD: 6.9}
& \makecell[c]{WER: 40.3\\
               CE/PQ: 6.42/7.04}
\\
\bottomrule
\end{tabular}
\end{table*}

\subsection{Performance on Zero-shot Tasks}

We further evaluate UniAudio~2.0 on more challenging zero-shot tasks to assess its generalization ability to unseen scenarios. Specifically, we first evaluate its text understanding capability on the MMLU benchmark~\cite{mmlu} in a zero-shot setting. Following~\cite{llama-omni}, we use the InstructS2S-Eval benchmark to evaluate speech conversation ability without task-specific examples. In addition, we design three new tasks: dysarthric speech recognition (DSR)~\cite{DSR}, speech-sound generation (Speech-S), and audio-prompt- and caption-guided speech generation (A-I-TTS). Table~\ref{tab:zeroshot_all} reports the results. We make the following observations. First, UniAudio~2.0 demonstrates strong text understanding ability, and the introduction of audio modalities does not significantly degrade text performance. Second, the model performs well on many unseen tasks, such as speech-to-text conversation and dysarthric speech recognition. Third, it exhibits promising generalization ability on newly designed tasks. For example, in speech-sound generation, the model is required to generate speech and sound events based on textual content and sound tags. Moreover, we observe that UniAudio~2.0 can follow instructions to leverage the timbre of an audio prompt and the style specified in a caption to synthesize the desired speech. Nevertheless, these tasks are not yet solved perfectly, and we summarize the limitations in Appendix~\ref{appendix:limitation}.

\subsection{Ablation Study}

\textbf{Influence of reasoning tokens.} We first investigate the impact of reasoning tokens. As shown in Tables~\ref{tab:ablation_llm_ppl_speech_music} and~\ref{tab:downstream_understanding}, removing reasoning tokens in both understanding and generation settings leads to worse performance than combining reasoning and reconstruction tokens. In Appendix~\ref{appendix:codec_exp}, we present additional ablation studies on ReasoningCodec, including the influence of the GRPO training loss (Appendix~\ref{appendix:codec_exp_grpo}), the effect of multi-expert semantic encoders (Appendix~\ref{appendix:codec_exp_mixture_semantic}), the effectiveness of FiLM (Appendix~\ref{appendix:codec_exp_film}), comparisons with previous semantic tokenizers (Appendix~\ref{appendix:codec_exp_with_other_tokenizer}), and the impact of classifier-free guidance (CFG) on reconstruction performance (Appendix~\ref{appendix:codec_exp_cfg}).

\textbf{Influence of multi-stage training.} We compare performance across different training stages. As shown in Table~\ref{tab:ablation_stage}, removing Stage~4 (the mid-training stage) degrades performance on several tasks, particularly those related to text capability. We attribute this degradation to the reduced text exposure and shorter context length during training, which weakens text understanding. Furthermore, without this mid-training stage, the model performs poorly on few-shot and zero-shot evaluations. These results suggest that increasing data and task diversity, for example by introducing auditory sentences, is crucial for improving generalization to unseen tasks.

\textbf{Influence of layer specialization.} We conduct experiments to evaluate the effectiveness of the proposed functional layer specialization. Specifically, we ablate the audio understanding and generation experts and train only the cross-modal expert using Stage~3. As shown in Table~\ref{tab:ablation_stage}, this modification leads to a significant performance degradation across multiple benchmarks. These results further validate the effectiveness of our unified autoregressive architecture. Moreover, our architectural modifications introduce no additional infrastructure complexity and remain fully compatible with tensor, pipeline, and context parallelism.

\textbf{Influence of model size.} We investigate the impact of the cross-modal expert's model size. Due to resource constraints, we use LLaMA~3.2~1B and LLaMA~3.2~3B as the cross-modal expert, respectively, and adopt the same training strategy for both variants. As shown in Table~\ref{tab:ablation_stage}, the 1B model exhibits a significant performance drop compared to the 3B model across multiple benchmarks. Appendix~\ref{appendix:zero-shot_1b_exp} further shows that its generalization ability on few-shot and zero-shot tasks is substantially weaker. These results indicate that model size is a critical factor in determining model capacity, especially for multi-task foundation models. Consistent with our findings, OpusLM~\cite{opuslm} also reports that model scale plays a crucial role in unified understanding and generation models.


\section{Conclusion}

In this work, we present a unified audio foundation model that supports both understanding and generation. We propose ReasoningCodec, which factorizes audio into reasoning and reconstruction tokens, and train UniAudio~2.0 using a unified autoregressive architecture together with a multi-stage, multi-task training strategy. Extensive experiments demonstrate strong and consistent performance on seen speech, sound, and music tasks, as well as promising few-shot and zero-shot generalization to unseen scenarios. 

Our ablation studies further indicate that scaling data and task diversity, as well as model capacity, plays a crucial role in improving generalization. In future work, we will continue to scale both model size and training data to further enhance the robustness and generalization ability of unified audio foundation models. The limitations of our approach are discussed in Appendix~\ref{appendix:limitation}.

\section{Impact Statement}
This work studies a unified audio foundation model for understanding and generation. While such models enable beneficial applications (e.g., creative assistance and human-computer interaction), they also introduce potential risks. \textbf{(1) misuse and harm}: Audio generation and voice conversion capabilities can be misused for impersonation, fraud, harassment, or the creation of unauthorized content. To mitigate these risks, we encourage responsible deployment practices, including clear user consent for voice cloning, identity verification in high-stakes settings, and downstream safeguards such as content provenance and detection where applicable. \textbf{(2) Copyright and content ownership}: Music and audio generation can reproduce or closely imitate styles present in training data, raising copyright and attribution concerns. We advise that generated content should not be used to infringe on copyrighted works, and that deployments should incorporate policy constraints and usage guidelines aligned with local regulations.

\nocite{langley00}

\bibliography{example_paper}
\bibliographystyle{icml2026}

\newpage
\appendix
\onecolumn

\section{ReasoningCodec} \label{appendix:reasoningcodec}
In this section, we provide additional details about the proposed ReasoningCodec. Figure \ref{fig:overview-codec} provides an overview of the ReasoningCodec framework.

\subsection{Model Structure of ReasoningCodec}
Table \ref{tab:codec-config} gives the details of ReasoningCodec configuration. 
\begin{table}[t]
  \centering
    \begin{tabular}{c|c}
    \toprule
           & ReasoningCodec \\
    \midrule 
    Input shape & (B, 1, N) \\
    \midrule
    \textbf{Reasoning Branch} \\
    Encoder modules  & Whisper \& Music Encoder \\
    Token Interleaving and Retrieval & w=5 \\
    Dimension of transformer encoder & 768 \\
    The number of transformer encoder & 4 \\
    Codebook size & 1024 \\
    VQ layers  & 8 \\
    \midrule
    \textbf{Reconstruction Branch} \\
    Encoder modules  & Whisper \& WavLM \& Music Encoder \\
    Codebook size & 1024 \\
    VQ layers  & 8 \\
    Semantic decoder layers & 4 convolutional layers \\
    \midrule
    \textbf{Flow-based Scalar Diffusion Decoder} \\
    The number of transformer decoder & 24 \\
    Dimension of transformer decoder & 768 \\
    Latent space dimension & 136 \\
    \bottomrule
    \end{tabular}
  \vspace{5pt}
  \caption{ReasoningCodec model backbone configurations}
  \label{tab:codec-config}
\end{table}

\subsection{The training details of ReasoningCodec} \label{appendix:codec_training_details}
The training of ReasoningCodec includes three stages: (1) supervised fine-tuning (SFT) on large-scale audio understanding tasks for the reasoning branch; (2) GRPO-style reinforcement learning to encourage detailed and grounded analytical descriptions and improve the perceptual reasoning capability of the reasoning branch; and (3) freezing the reasoning branch and training the reconstruction branch with a flow-based decoder. Table \ref{tab:codec_training_stages} presents the training data and configurations for each stage. We build the mixture of audio reasoning data \footnote{We use the term reasoning to denote fine-grained, perceptual analyses and inference cues grounded in audio (e.g., events, rhythm, timbre, and scene context), rather than symbolic multi-step logical deduction.} based on Qwen3-Omni \cite{qwen-omni} and Gemini 2.5 Pro. Boxes \ref{box:speech_case}--\ref{box:music_case} show examples. Below, we describe each stage in detail.

\subsubsection{Stage 1: SFT for Reasoning Branch}
In this stage, we follow SALMONN \cite{salmonn} and use multiple audio understanding tasks as training objectives. During training, we only update the VQ modules and LoRA parameters to preserve pretrained linguistic knowledge. For different tasks, we prepare multiple prompts and randomly sample one prompt per instance.

\subsubsection{Stage 2: GRPO for Reasoning Branch}
In this stage, we follow common GRPO practice \cite{shao2024deepseekmath,r1-aqa} and design an accuracy-based reward to train the model. Specifically, we classify audio understanding tasks into two categories: (1) rule-based verifiable tasks, such as ASR and audio classification, where we use WER and label accuracy as rewards; and (2) tasks that are difficult to verify automatically, such as fine-grained audio analysis and interpretation. For the latter, we introduce an LLM-based judge to score each rollout. We use LLaMA~3.1-Instruct~8B as the judge and ask it to evaluate the gap between the rollout and the ground-truth answer. Table \ref{tab:grpo_hparams} presents the hyper-parameters used for GRPO training.
Given an input query $q$, the model first samples $G$ distinct outputs $\{o_1, o_2, \dots, o_G\}$.
Each output is evaluated by our reward model $R$, which assigns a scalar reward $r_i$ to each sample.
Based on these rewards, the relative advantage for each output can be computed as shown in Equation \ref{formula:grpo_1}.
The full GRPO objective is summarized in the Equation \ref{formula:grpo_2}.
\begin{align}
\label{formula:grpo_1}
\hat{A}_{i,t} &=
\frac{r(q, o_i) - \mathrm{mean}\{r(q,o_1), \ldots, r(q,o_G)\}}
     {\mathrm{std}\{r(q,o_1), \ldots, r(q,o_G)\}}
\end{align}

\begin{align}
\label{formula:grpo_2}
\mathcal{L}_{\text{GRPO},i,t}
&=
\min\Bigg[
\frac{\pi_{\theta}(o_{i,t} \mid q, o_{i,<t})}
     {\pi_{\theta_{\text{old}}}(o_{i,t} \mid q, o_{i,<t})}
     \, \hat{A}_{i,t}, \nonumber\\[-2pt]
&\qquad
\mathrm{clip}\!\left(
\frac{\pi_{\theta}(o_{i,t} \mid q, o_{i,<t})}
     {\pi_{\theta_{\text{old}}}(o_{i,t} \mid q, o_{i,<t})},
     1-\epsilon,\,
     1+\epsilon
\right)\hat{A}_{i,t}
\Bigg]
\end{align}
where $\epsilon$ denotes the PPO clipping range.

\subsubsection{Stage~3: Waveform Reconstruction with a Flow-based Diffusion Decoder}
\label{appendix:codec_training_stage3}

In this stage, we freeze the reasoning branch and train the reconstruction branch to recover waveforms from discrete tokens. Specifically, we first employ a multi-expert semantic encoder to extract semantically rich features from speech, music, and sound signals. These features are then quantized into discrete tokens using three parallel VQ groups, with a 1:1:6 allocation of VQ layers across the three groups. We assign six VQ layers to the last group to encode finer-grained acoustic details for high-fidelity reconstruction. 

Overall, the three groups produce eight discrete tokens per frame, which we refer to as \emph{reconstruction tokens}. Following SimpleSpeech~\cite{simplespeech2}, we further construct a flow-based scalar diffusion decoder to reconstruct waveforms from these tokens.

We do not directly condition waveform reconstruction on reasoning tokens for two main reasons. First, reconstruction tokens already capture the essential information provided by reasoning tokens. Second, explicit conditioning would significantly increase the sequence length of the diffusion transformer. In preliminary experiments, upsampling reasoning tokens and concatenating them with reconstruction tokens resulted in only marginal improvements. Instead, we adopt FiLM~\cite{perez2018film} to inject reasoning-token information into the multi-expert semantic encoder. Empirically, this design introduces no noticeable degradation in reconstruction quality.

For SQ-Codec, we follow the training protocol of SimpleSpeech~\cite{simplespeech} and set the latent dimension to 136. The model is trained on the same dataset used in Stage~3, as summarized in Table~\ref{tab:codec_training_stages}.

\subsubsection{Implementation Details}
We train ReasoningCodec in multiple stages. For each stage, we train the model on 8 NVIDIA A100 GPUs. The learning rate is set to $1e-4$ with a cosine annealing schedule. \\

\begin{table}[t]
\centering
\caption{Details of the hyper-parameters used for GRPO training.}
\label{tab:grpo_hparams}
\setlength{\tabcolsep}{8pt}
\renewcommand{\arraystretch}{1.08}
\begin{tabular}{@{}l c@{}}
\toprule
\textbf{Setting} & \textbf{Value} \\
\midrule
Batch Size per Device               & 1 \\
Gradient Accumulation Steps         & 2 \\
Learning Rate                       & $1 \times 10^{-6}$ \\
Temperature                         & 1.0 \\
Maximum Response Length             & 2048 \\
Number of rollouts                 & 8 \\
Kullback-Leibler Coefficient        & 0.04 \\
\bottomrule
\end{tabular}
\end{table}

\begin{table}[t]
\centering
\caption{Training stages and configurations of ReasoningCodec. The first two stages train the reasoning branch, and the last stage trains the reconstruction branch and decoder while freezing the reasoning branch.}
\label{tab:codec_training_stages}
\setlength{\tabcolsep}{6pt}
\renewcommand{\arraystretch}{1.25}
\begin{tabular}{c >{\centering\arraybackslash}p{0.56\textwidth} >{\centering\arraybackslash}p{0.24\textwidth}}
\toprule
\textbf{Setting} & \textbf{Training data} & \textbf{Other configuration} \\
\midrule
Stage 1 &
WenetSpeech~\cite{wenetspeech}, MLS~\cite{pratap2020mls}, AudioSet~\cite{audioset}, IEMOCAP \cite{iemocap}, AudioCaps \cite{audiocaps}, WavCaps \cite{mei2023wavcaps}, LibriSpeech \cite{librispeech}, TUT Acoustic scenes 2017 \cite{tut2017} LP-MusicCaps~\cite{lp-musiccaps}, mixture of audio reasoning data &
learning rate = 2e-4 \newline
LoRA rank = 64 \\
\midrule
Stage 2 &
LibriSpeech \cite{librispeech}, AudioCaps \cite{audiocaps}, small-set of LP-MusicCaps \cite{lp-musiccaps}, IEMOCAP \cite{iemocap}, AudioCaps \cite{audiocaps}, WavCaps \cite{mei2023wavcaps}, LibriSpeech \cite{librispeech}, TUT Acoustic scenes 2017 \cite{tut2017}, mixture of audio reasoning data &
learning rate = 1e-6 \\
\midrule
Stage 3 &
Million Song \cite{msd}, MLS \cite{pratap2020mls}, AudioSet \cite{audioset} &
learning rate = 2e-4 \\
\bottomrule
\end{tabular}
\end{table}

\begin{figure*}[t]
    \centering
    \includegraphics[width=0.8\textwidth, height=0.4\textwidth]{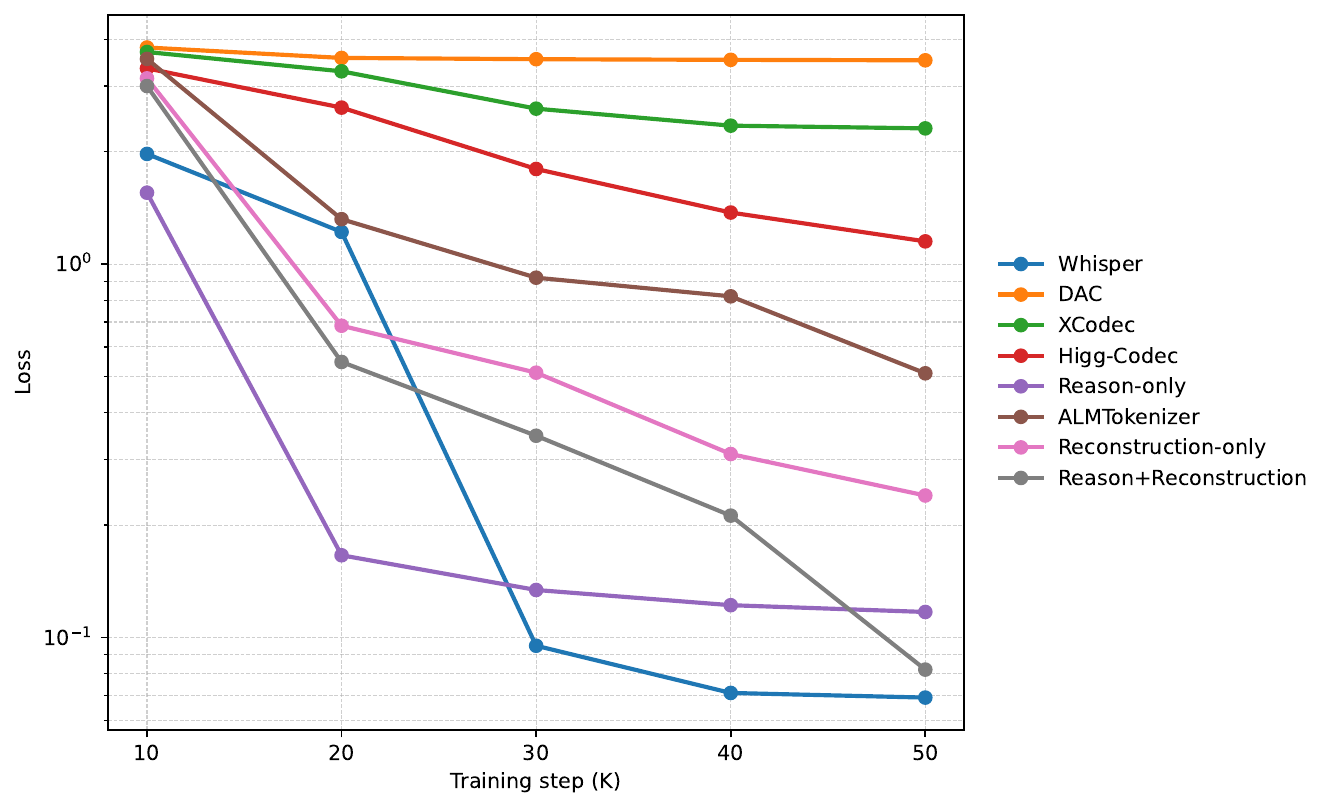}
    \caption{The training loss of different audio tokenizer for understanding tasks.}
    \label{fig:codec_understanding_loss}
\end{figure*}

\begin{table*}[t]
\centering
\setlength{\tabcolsep}{5pt}
\caption{Ablation study for the effectiveness of FiLM. Similarly, we also use the PPL across codebooks as the metric.}
\begin{tabular}{ccccccccccc}
\hline
\multirow{2}{*}{Model} &
\multicolumn{5}{c}{Speech} &
\multicolumn{5}{c}{Music} \\
\cmidrule(lr){2-6}\cmidrule(lr){7-11}
& VQ1 & VQ2 & VQ3 & VQ4 & Avg
& VQ1 & VQ2 & VQ3 & VQ4 & Avg \\
\hline
Reason + Reconstruction & 6.92 & 11.37& 19.42 & 55.36 & {23.77} & 7.02 & 15.55& 15.23 & 42.5  & {20.1} \\
W/O FiLM & 7.22 & 12.92& 23.45 & 60.12 & {25.9} & 7.98 & 17.23 & 19.67 & 54.6  & {24.9} \\
\hline
\end{tabular}
\label{tab:appendix_film}
\end{table*}

\begin{table*}[t]
\centering
\setlength{\tabcolsep}{2pt}
\caption{The influence of FiLM module for the Reconstruction performance. PESQ includes two versions: WB and NB. AudioBox Score includes CE, CU, PC, and PQ.}
\begin{tabular}{cccccccccc}
\hline
\multirow{2}{*}{\textbf{Model}} &
\multicolumn{5}{c}{\textbf{Speech}} &
\multicolumn{2}{c}{\textbf{Sound}} &
\multicolumn{2}{c}{\textbf{Music}} \\
\cmidrule(lr){2-6}\cmidrule(lr){7-8}\cmidrule(lr){9-10}
& \textit{PESQ} & \textit{STOI} & \textit{UT-MOS} & \textit{VISQOL} & \textit{SIM}
& \textit{VISQOL} & \textit{AudioBox Score}
& \textit{VISQOL} & \textit{AudioBox Score} \\
\hline
ReasonCodec          & 2.36/2.93 & 0.85 & 3.91 & 3.94 & 0.97 & 3.10 & 4.12/5.06/3.58/5.96 & 4.03 & 7.51/7.68/6.12/7.87 \\
W/O FiLM          & 2.39/2.92 & 0.85 & 3.90 & 3.92 & 0.97 & 3.14 & 4.21/5.12/3.49/5.92 & 4.01 & 7.48/7.62/6.14/7.78 \\
\hline
\end{tabular}
\label{tab:ablation_codec_recon}
\end{table*}

\subsection{Experiments} \label{appendix:codec_exp}
In this part, we include additional more experiments for ReasoningCodec, focusing on the effectiveness of GRPO, why choose multi-expert semantic features, the effectiveness of FiLM, the subjective evaluation for audio tokenizer, why the reasoning tokens are important, the comparison between ReasoningCodec and other semantic tokenizers, and the influence of Classifier-free guidance for reconstruction performance.

\subsubsection{The effectiveness of GRPO for the reasoning branch training} \label{appendix:codec_exp_grpo}
To better understand the influence of GRPO training, we use the same audio understanding test set to evaluate the reasoning branch on ASR and audio classification tasks.
As Table~\ref{tab:appendix_sft_grpo_ablation} shows, GRPO further improves performance on audio understanding tasks and also improves the quality of detailed audio reasoning analysis. Overall, GRPO brings a significant improvement.

\begin{table}[t]
\centering
\caption{Ablation results of SFT and GRPO on understanding tasks.  Audio CLS denotes the audio classification tasks. For Audio Reasoning, we report GPT-score in terms of relevance and fluency.}
\label{tab:appendix_sft_grpo_ablation}
\setlength{\tabcolsep}{9pt}
\renewcommand{\arraystretch}{1.15}
\begin{tabular}{cccc cc}
\toprule
\multirow{2}{*}{\textbf{Setting}} & \multirow{2}{*}{\textbf{ASR}} & \multirow{2}{*}{\textbf{Audio CLS}} &
\multicolumn{2}{c}{\textbf{Audio Reasoning (GPT-score)}} \\
\cmidrule(lr){4-5}
& & & \textbf{Relevance} & \textbf{Fluency} \\
\midrule
SFT       & 10.68 & 36.4 & 5.2 & 7.8 \\
SFT+GRPO  & \textbf{7.64} & \textbf{40.2} & \textbf{5.7} & \textbf{8.4} \\
\bottomrule
\end{tabular}
\end{table}

\subsubsection{The effectiveness of multi-expert semantic features}
 \label{appendix:codec_exp_mixture_semantic}
Different audio modalities may emphasize different types of semantic information. We therefore use multiple expert encoders to extract modality-specific semantic features and apply separate VQ heads to quantize them.
To demonstrate the effectiveness of this strategy, we build two baselines: (1) concatenating all semantic features and applying query-based quantization; and (2) concatenating all semantic features and applying standard RVQ. Table \ref{tab:quant_compare} shows the results. From the reconstruction perspective, query-based quantization performs better, while group-wise VQ yields better token modeling performance (semantic information), especially for music. We attribute this to using separate VQ layers for different semantic features.
Although query-based quantization can further improve reconstruction, it requires multiple transformer encoders. Considering the additional inference cost, we do not apply query-based quantization in the reconstruction branch. 

\subsubsection{The effectiveness of FiLM}  \label{appendix:codec_exp_film}
To explore whether using FiLM \cite{perez2018film} to connect both reasoning branch and reconstruction branch is useful, we design an ablation study, as shown in Table \ref{tab:ablation_llm_ppl_speech_music}. We can see that using FiLM to build the connection between reasoning tokens and reconstruction tokens is effective. Furthermore, as Table \ref{tab:ablation_codec_recon} shows, introducing the FiLM module does not influence the reconstruction performance.

\subsubsection{The subjective evaluation for audio tokenizer} \label{appendix:mos_eval}
For the subjective evaluation, we follow previous works \cite{almtokenizer,kreuk2022audiogen} conduct the MUSHRA test. 
Table \ref{tab:mos} shows the subjective evaluation. We can see that our proposed ReasoningCodec obtains best performance on both speech, sound, and music reconstruction.

\subsubsection{Why Reasoning tokens are important for both understanding and generation tasks?} \label{appendix:codec_why_reason}
As the results in Table~\ref{tab:ablation_llm_ppl_speech_music} and Table~\ref{tab:downstream_understanding} show, introducing \textit{Reasoning tokens} consistently improves understanding performance and reduces the modeling difficulty (e.g., lower PPL) of reconstruction tokens. We attribute the gains to two complementary factors.

\textbf{(1) A language-aligned bottleneck that filters task-irrelevant acoustic details.}
Reasoning tokens are explicitly designed to align with the latent space of the text LLM, thus encouraging the representation to retain language- and reasoning-relevant content (e.g., lexical/semantic cues, high-level events) while discarding irrelevant factors (e.g., recording condition, acoustic noisy) that are less useful for understanding. This results in a more learnable target for the LLM and faster convergence. As shown in Figure~\ref{fig:codec_understanding_loss}, the training loss on reasoning tokens decreases rapidly, suggesting that the LLM can efficiently capture the structure of these tokens under the autoregressive objective.

\textbf{(2) A shared intermediate representation that benefits both understanding and generation.}
From the understanding perspective, reasoning tokens serve as an intermediate feature layer that is directly optimized for semantic abstraction, providing a compact and discriminative signal for downstream tasks (ASR, classification, and reasoning-style evaluations). Compared with purely acoustic tokens or reconstruction tokens, this abstraction has better alignment with the text LLM.

From the generation perspective, reasoning tokens also act as a ``planning'' role \footnote{In the context of LLMs, this can be viewed as a latent reasoning process: the model first produces a high-level plan/thought, which then helps predict the final output.}: they summarize high-level intent and semantic content, which simplifies the subsequent prediction of fine-grained reconstruction tokens. Concretely, conditioning the generation process on reasoning tokens reduces long-range uncertainty and stabilizes autoregressive decoding, leading to lower reconstruction-token perplexity (Table~\ref{tab:ablation_llm_ppl_speech_music}) and improved controllability (e.g., better adherence to captions/instructions). Overall, reasoning tokens bridge text-aligned semantics and audio realizations, enabling a single autoregressive model to scale to diverse understanding and generation tasks with improved efficiency and generalization.
\paragraph{Information-theoretic view.}
From an information-theoretic perspective, reasoning tokens \(R\) act as a compact intermediate variable that reduces the uncertainty of reconstruction-token generation.
Let \(X\) denote the conditioning input (e.g., caption/instruction) and \(S_{1:T}\) the reconstruction token sequence.
For an autoregressive model, the optimal negative log-likelihood (NLL) decomposes as
\[
\mathcal{L}_{\text{sem}}^{\star}
= \sum_{t=1}^{T} H\!\left(S_t \mid X, S_{<t}\right).
\]
Introducing reasoning tokens yields
\[
\mathcal{L}_{\text{sem}}^{\prime\star}
= \sum_{t=1}^{T} H\!\left(S_t \mid X, R, S_{<t}\right),
\]
and conditioning can only reduce entropy:
\[
H\!\left(S_t \mid X, R, S_{<t}\right)
\le
H\!\left(S_t \mid X, S_{<t}\right).
\]
The reduction equals the conditional mutual information:
\[
H\!\left(S_t \mid X, S_{<t}\right)
-
H\!\left(S_t \mid X, R, S_{<t}\right)
=
I\!\left(S_t; R \mid X, S_{<t}\right),
\]
which directly explains the lower reconstruction-token perplexity.
Moreover, marginalizing over \(R\) gives a mixture decomposition,
\[
p\!\left(S_{1:T}\mid X\right) = \sum_{R} p\!\left(R\mid X\right)\, p\!\left(S_{1:T}\mid X,R\right),
\]
where \(R\) selects a high-level ``plan/reasoning'' stage, mitigating long-range multimodality and stabilizing autoregressive decoding.

\begin{figure*}[t]
    \centering
    \includegraphics[width=\textwidth]{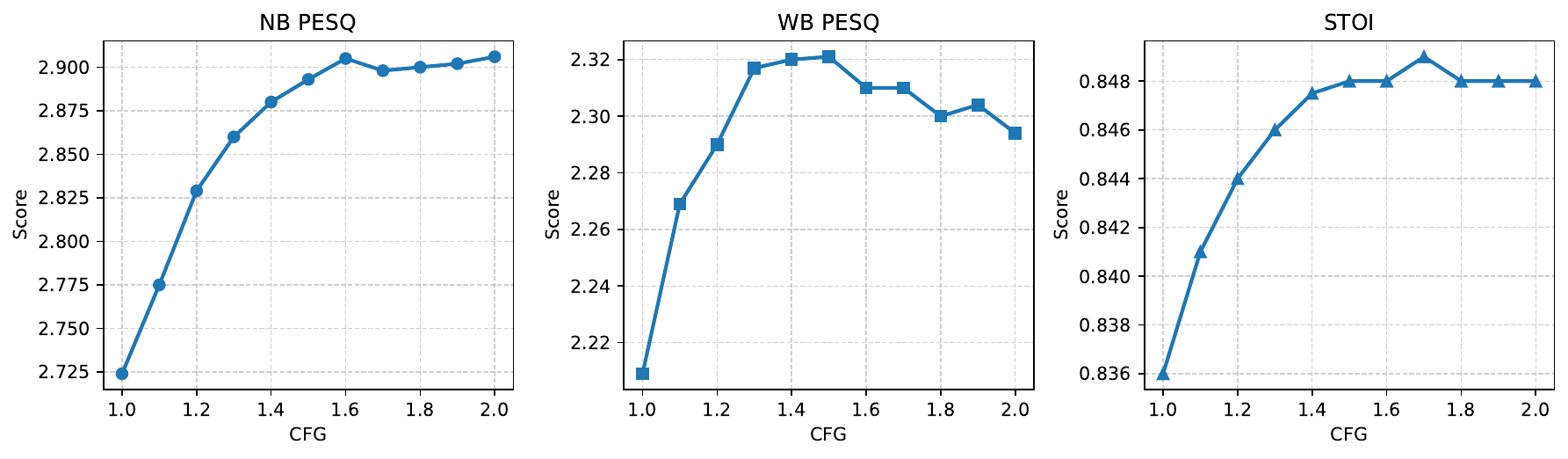}
    \caption{The influence of CFG for reconstruction performance.}
    \label{fig:cfg_val}
\end{figure*}

\subsubsection{Compared to previous semantic tokenizers}  \label{appendix:codec_exp_with_other_tokenizer}
To obtain the reasoning tokens, we introduce a multi-task audio understanding objective. Similar ideas of injecting supervision into a tokenizer/encoder have also been explored in CosyVoice~3~\cite{cosyvoice3} and USTokenizer~\cite{wang2025dualspeechlm}. Nevertheless, our target and design are fundamentally different from prior works in several key aspects. \\
\textbf{(1) Alignment to text LLM.} CosyVoice~3 inserts an FSQ layer into an intermediate encoder layer and updates the entire encoder during training. As a result, its learned semantic tokens are not explicitly constrained to align with the latent space of a text LLM. In contrast, we \emph{freeze} the audio encoder and learn only a lightweight linear projection that maps the reasoning tokens into the latent space of the text LLM, enabling direct and stable cross-modal alignment. \\
\textbf{(2) Modality coverage.} Prior work mainly targets the speech domain, whereas our model is designed for \emph{unified} understanding across speech, general sounds, and music. \\
\textbf{(3) Temporal granularity.} Previous approaches typically adopt finer-grained discrete representations (e.g., $\sim$25\,Hz). Our reasoning tokens operate at 5\,Hz, yielding a much more compact representation that is better suited for high-level latent reasoning/planning rather than dense acoustic content modeling. \\
\textbf{(4) Richer understanding targets.} Beyond task-level supervision, we incorporate more detailed understanding signals, encouraging the model to capture fine-grained audio attributes and to form an explicit reasoning process over audio details.

Furthermore, we apply the same training protocol to an alternative tokenizer (CosyVoice~3 tokenizer) on the same set of understanding tasks. As shown in Table~\ref{tab:codec_ablation_asr_er}, its performance is consistently worse than that of our reasoning tokens, validating the advantage of our design.

\subsubsection{The influence of classifier-free guidance and decoding steps}  \label{appendix:codec_exp_cfg}
In this part, we conduct experiments to explore the influence of  classifier-free guidance (CFG) and the diffusion steps for audio reconstruction. Figure \ref{fig:cfg_val} shows the relationship between speech reconstruction and CFG parameter. Based on this results, we default to use CFG = 1.5 for all of experiments. We also find the minimum diffusion step is 10 steps for speech and music. But for the sound data, we recommend to use 25 steps.

\begin{table}[t]
\centering
\caption{Ablation study of different semantic tokenizer. We report the ASR and emotion recognition tasks.}
\label{tab:codec_ablation_asr_er}
\setlength{\tabcolsep}{10pt}
\renewcommand{\arraystretch}{1.15}
\begin{tabular}{lcc}
\toprule
\textbf{Model} & \textbf{ASR} & \textbf{ER} \\
\midrule
CosyVoice3 \cite{cosyvoice3} & 32.3 & 34.5 \\
Ours (reasoning token) & \textbf{10.1} & \textbf{50.2} \\
\bottomrule
\end{tabular}
\end{table}

\begin{table}[t]
\centering
\small
\caption{Comparison of different quantization strategies. For each strategy, we control the same VQ layers (8 VQ layers), training data, and decoder.}
\label{tab:quant_compare}
\setlength{\tabcolsep}{6pt}
\renewcommand{\arraystretch}{1.12}
\begin{tabular}{ccccccccc}
\toprule
\textbf{Model} & \textbf{PESQ (WB)} & \textbf{PESQ (NB)} & \textbf{STOI} & \textbf{UT-MOS} & \textbf{ViSQOL} & \textbf{SIM} & $\bm{\mathrm{PPL}}_{\text{speech}}$ & \textbf{$\bm{\mathrm{PPL}}_{\mathrm{music}}$} \\
\midrule
RVQ & 2.12 & 2.76 & 0.83 & 3.84 & 3.78 & 0.95 & 32.8 & 29.2 \\
Query-based & \textbf{2.54} & \textbf{3.07} & \textbf{0.85} & 3.89 & 3.88 & \textbf{0.97} & 26.3 & 24.4 \\
Group-wise VQ & 2.36 & 2.93 & \textbf{0.85} & \textbf{3.91} & \textbf{3.94} & \textbf{0.97} & \textbf{23.77} & \textbf{20.1} \\
\bottomrule
\end{tabular}
\end{table}

\subsection{Audio Tokenizer Baselines}
To make a fair comparison, we classify audio tokenizers into two types: (1) speech-based tokenizers trained on speech datasets, and (2) audio-based tokenizers trained on speech, sound, and music datasets. In this study, we mainly compare against audio-based tokenizers trained on speech, sound, and music datasets. Below, we list our chosen baselines:

(1) Encodec \cite{encodec}, a SOTA audio codec model trained on large-scale speech, sound, and music datasets. We use the official open-sourced 24~kHz version. The frame rate is 75~Hz. 

(2) DAC-Codec \cite{dac}, which offers very high reconstruction performance. It is trained on large-scale speech, sound, and music datasets. The official open-sourced 24 kHz version is used. The sampling rate is 16 kHz, the frame-rate is 50hz. 

(3) MimiCodec \cite{moshi}, a SOTA low-bitrate speech codec model trained on a large-scale speech dataset. The sampling rate is 24 kHz, the frame-rate is 12.5hz. 

(4) X-Codec \cite{xcodec}, a semantic-rich audio codec model trained on a large-scale speech, sound, and music datasets. The sampling rate is 16k Hz, the frame-rate is 50hz. 

(5) Higgs-Audio-tokenizer\footnote{https://github.com/boson-ai/higgs-audio}, a SOTA audio codec model trained on a large-scale speech, sound, and music dataset. The sampling rate is 24k Hz, the frame-rate is 25hz. 

To make a fair comparison, for Encodec, DAC-Codec, X-Codec, and Higgs-Audio-tokenizer, we control the bitrate by dropping some RVQ layers during inference.

\begin{tcolorbox}[
  colback=white,
  colframe=black!60,
  boxrule=0.8pt,
  arc=3mm,
  left=6pt,right=6pt,top=6pt,bottom=6pt,
  title=\textbf{Speech Reasoning Case},
  fonttitle=\bfseries,
  coltitle=white,
  colbacktitle=black!70,
  listing only,
  listing options={
    basicstyle=\ttfamily\small,
    breaklines=true,
    breakatwhitespace=true,
    columns=fullflexible
  }
]
\label{box:speech_case}
The audio begins with a clear, albeit slightly muffled, sound.
The speaker's gender is discernible as male. The language is English.
There is only one speaker. The recording environment seems to be indoors, possibly a quiet room,
but with some ambient background noise, suggesting it's not a professional studio.
The speaker exhibits a moderate speech rate. There are no significant pauses indicating hesitation or thought,
suggesting a natural flow. The pitch variation is relatively consistent, without dramatic shifts,
implying a neutral to slightly conversational prosody. The pronunciation is generally fluent and clear,
with no noticeable impediments.
The audio contains the spoken phrase '00:03'. This phrase is a time indicator.
The emotional tendency is neutral, and the tone is informative or declarative.
\end{tcolorbox}

\begin{tcolorbox}[
  colback=white,
  colframe=black!60,
  boxrule=0.8pt,
  arc=3mm,
  left=6pt,right=6pt,top=6pt,bottom=6pt,
  title=\textbf{Sound Reasoning Case},
  fonttitle=\bfseries,
  coltitle=white,
  colbacktitle=black!70,
  listing only,
  listing options={
    basicstyle=\ttfamily\small,
    breaklines=true,
    breakatwhitespace=true,
    columns=fullflexible
  }
]
\label{box:sound_case}
The audio consists of a distinct, rhythmic pattern of impacts. These impacts are predominantly in the mid-frequency range and suggest a mechanical or physical action occurring outdoors or in a large space. There is a clear periodic repetition. The primary sound event is a repeating impact, sounding like a hammer or a similar percussive tool. The first instance occurs around 00:00:01 and repeats approximately every second, with a sharp, distinct onset. The duration of each impact is very short, less than half a second. The spectral content is broadband but with a strong emphasis on mid to high frequencies due to the impact. There's a slight sense of reverberation, suggesting the sound is not in a completely enclosed space but perhaps outdoors or in a large hall. The sound appears to be localized to a central or slightly directional source. The repetitive striking sound strongly suggests construction work, hammering, or perhaps pile driving. The consistent rhythm and sharp impacts indicate a deliberate and ongoing activity. The emotional connotation is neutral to slightly industrious, not alarming. The source interaction is minimal, with the sound of the impact itself being the dominant element, potentially masking very quiet background noises. The action appears to be a single, repeating task.
\end{tcolorbox}

\begin{tcolorbox}[
  colback=white,
  colframe=black!60,
  boxrule=0.8pt,
  arc=3mm,
  left=6pt,right=6pt,top=6pt,bottom=6pt,
  title=\textbf{Music Reasoning Case},
  fonttitle=\bfseries,
  coltitle=white,
  colbacktitle=black!70,
  listing only,
  listing options={
    basicstyle=\ttfamily\small,
    breaklines=true,
    breakatwhitespace=true,
    columns=fullflexible
  }
]
\label{box:music_case}
The audio has a moderate to high fidelity, with clear instrumentation and vocals. The genre appears to be rock or hard rock, characterized by distorted electric guitars, a prominent drum beat, and aggressive vocals. The main instruments are electric guitar, bass guitar, and drums, with male vocals. The tempo is moderately fast, around 120 BPM, with a consistent and driving beat. The key appears to be A minor, and the chord progression is likely based on standard rock progressions, possibly with power chords. The meter is 4/4 time, and the rhythmic complexity is moderate, with a strong backbeat. The melody is delivered by the vocals and guitar, with a relatively narrow range in the vocals and more varied melodic lines in the guitar riffs. The timbre is dominated by distorted guitars, a solid bass tone, and punchy drums. Dynamics are generally loud, with some variation in intensity. The mood is energetic, aggressive, and defiant. The lyrics, though partially obscured by the music, seem to convey a sense of rebellion or challenging authority ('smash the machine', 'ain't no stopping'). The expressive techniques include distorted guitar tones, heavy drumming with prominent cymbal work, and powerful, somewhat raw vocals. A potential hook can be identified in the main guitar riff and vocal chorus, which is memorable and driving. The structure appears to follow a typical verse-chorus rock song format, likely with an intro, verses, choruses, and potentially a bridge or guitar solo (though not fully audible in this short clip).
\end{tcolorbox}

\section{UniAudio 2.0} \label{appendix:uniaudio2}

\subsection{Model Configuration}
We list the detailed configuration of UniAudio 2.0 in Table \ref{tab:uniaudio2_arch_config}. 

\begin{table}[t]
\centering
\caption{UniAudio~2.0 architecture configuration.}
\label{tab:uniaudio2_arch_config}
\setlength{\tabcolsep}{6pt}
\renewcommand{\arraystretch}{1.12}
\begin{tabular}{cc}
\toprule
\textbf{Module} & \textbf{Configuration} \\
\midrule
Backbone & Init from \texttt{LLaMA-3.2-3B} \\
\midrule

Audio understanding experts &
$\#\text{layers}=3$;\;
$ d_{\text{model}}=\texttt{3072}$;\;
$ \#\text{heads}=\texttt{24}$;\;
$ T=\texttt{[2048]}$ \\

Cross-modal experts &
$\#\text{layers}=28$;\;
$ d_{\text{model}}=\texttt{3072}$;\;
$ \#\text{heads}=\texttt{24}$;\;
$ T=\texttt{[2048]}$ \\

Audio generation experts  &
$\#\text{layers}=2$;\;
$ d_{\text{model}}=\texttt{3072}$;\;
$ \#\text{heads}=\texttt{24}$;\;
$ T=\texttt{[2048]}$ \\

Audio Local Decoder&
$\#\text{layers}=4$;\;
$ d_{\text{model}}=\texttt{2048}$;\;
$ \#\text{heads}=\texttt{32}$;\;
$ T=\texttt{[8]}$ \\

\midrule
Total layers & $L=37$ \\
\midrule
Text tokenizer & \texttt{LLaMA3.2 tokenizer} \\
Audio tokenizer & \texttt{ReasoningCodec} \\
Audio sample rate & \texttt{[24 kHz]} \\
Reasoning token rate & \texttt{[5 Hz]} \\
Reconstruction token rate & \texttt{[12.5 Hz]} \\
\bottomrule
\end{tabular}
\end{table}

\subsection{Training Data Details} \label{appendix:uniaudio_training_data}

We organize the training corpus into several data types, each corresponding to a class of tasks supported by UniAudio 2.0. The training data includes the following data types:

\paragraph{Text-only data}
To maintain and stabilize the language modeling capability of UniAudio~2.0, we incorporate large-scale text-only data during pre-training. Specifically, we use approximately $100$B tokens from the high-quality annealed corpus of OLMo~3~\citep{olmo2025olmo}. This data is used to maintain the text capability of the original LLM when we introduce audio-modality data. 

\paragraph{Audio-only data}
To expose the model to diverse acoustic patterns beyond speech, we use large-scale unlabeled audio data from LAION-Audio-300M.\footnote{\url{https://huggingface.co/datasets/laion/LAION-Audio-300M}} This dataset covers speech, environmental sounds, and music.

\paragraph{Speech-transcription data}
Paired speech and transcription data are used to construct automatic speech recognition (ASR) and text-to-speech (TTS) tasks. We collect such data from multiple sources, including Emilia~\citep{emilialarge}, YODAS-English~\citep{yodas}, WenetSpeech~\citep{wenetspeech}, and WenetSpeech-Yue~\citep{wenetspeech-yue}.

\paragraph{Speech-caption-transcription data}
To support speech captioning and instruction-following TTS tasks, we use datasets that provide speech, transcription, and descriptive captions. In particular, we adopt CapSpeech-MLS dataset ~\citep{capspeech}, following prior work on instruction-based speech generation~\citep{yang2023instructtts}.

\paragraph{Audio-caption data}
Audio-caption pairs enable text-to-audio and text-to-music generation. We aggregate such data from multiple sources, including WavCaps \cite{mei2023wavcaps}, AudioSet \cite{audioset}, the Million Song Dataset (MSD) \cite{msd}, and YouTube-8M \cite{youtube-8m}. 

\paragraph{Lyric-song data}
To model lyric recognition and lyric-to-song generation, we construct lyric--song pairs by following the preprocessing pipeline of SongGen~\citep{liu2025songgen} on the MSD dataset, Free Music Archive \cite{fma_dataset} and MTG-Jamendo Dataset \cite{bogdanov2019mtg}.

\paragraph{Auditory sentences} \label{appendix:auditory}
To further improve compositional generalization and long-context reasoning over audio, we introduce the concept of \emph{auditory sentences}, inspired by ~\citep{bai2023sequential}. 
An auditory sentence is a long-context training sequence composed of multiple, related segments (audio and/or text), designed to encourage the model to reason over compositional structures and cross-segment dependencies. We construct such sequences using several strategies: 

(1) segmenting long-form audio such as speech conversations, environmental recordings, or songs: we split each recording into 2--8 segments (ensuring the token sequence length does not exceed 2048). We choose these long-form audio samples from LAION-Audio-300M, LibriLight \cite{librilight}, WavCaps, MSD, DailyTalk \cite{dailytalk}, and Expresso \cite{expresso}. 

(2) interleaving speech and text segments: to build this data, we use the MLS \cite{pratap2020mls} and LibriSpeech \cite{librispeech} datasets. We randomly choose several speech-transcription pairs from the same speaker to construct such sentences. We fix the order of speech and text in each sentence to keep the input format consistent.

(3) interleaving audio/music and captions: similarly, we use the pairs from AudioSet, WavCaps, AudioCaps, and MSD to build auditory sentences with alternating audio and caption segments.

(4) Building the mixture-clean triples: We randomly choose two audio samples ($a$ and $b$) and build a mixture audio $c$. We then form an auditory sentence such as \{$a, b, c$\} or \{$c, a, b$\} or \{$c, b, a$\} to encourage mixture reasoning. We can also concatenate multiple such triples, e.g., \{$a_1,b_1,c_1,a_2,b_2,c_2,\ldots$.\} For both $a$ and $b$, they can be speech, sound, or music samples from a pre-defined dataset $A$. The dataset $A$ consists of speech data from MLS, LibriSpeech, and WenetSpeech, sound data from AudioSet and WavCaps, and music data from MSD. Furthermore, we use demucs \cite{demucs} to separate music into vocals and accompaniment to improve data diversity.

(5) Building semantic-consistent but acoustically varied pairs: we follow InstructSpeech \cite{instructspeech} and use a TTS model to construct multiple speech samples that have the same text content but different pitch, volume, speed, and emotion. This encourages the model to distinguish fine-grained acoustic attributes from semantic content.

\subsection{The details of multiple stage training} \label{appendix:uniaudio2_4_stage_training}
\begin{figure*}[t]
    \centering
    \includegraphics{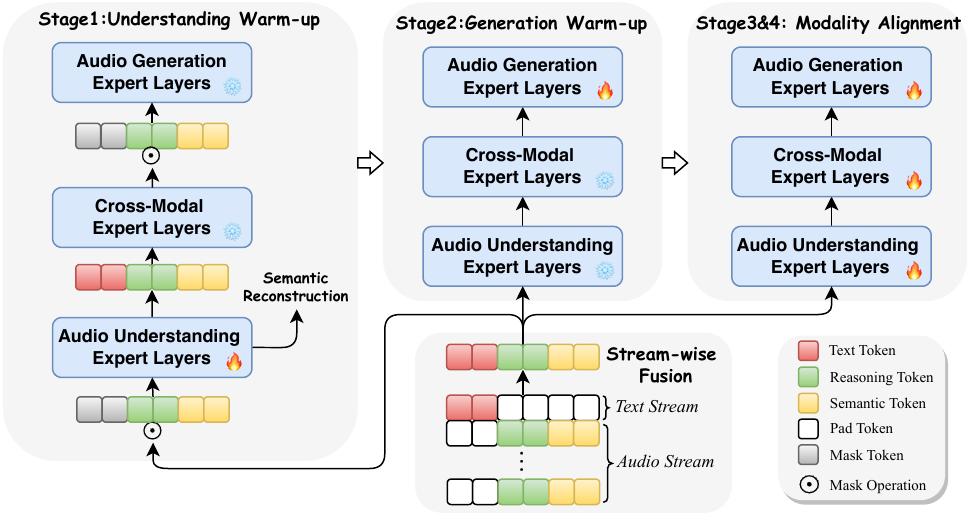}
    \caption{Overview of the multi-stage pre-training strategy of UniAudio~2.0. 
The model is trained in a progressive manner, including an understanding warm-up stage, a generation warm-up stage, and a modality alignment stage, with different expert layers gradually unfrozen. Stream-wise fusion is employed to merge text and audio streams under a unified autoregressive objective. 
For simplicity, the audio generation expert layers include both the generation expert and the local decoder.
}
    \label{fig:overview-training-stage}
\end{figure*}


\begin{table*}[t]
\centering
\caption{Four-stage training recipe of UniAudio~2.0. ``Trainable'' indicates the modules updated at each stage, and ``Ctx'' denotes the maximum context length used during training. LR denotes the learning rate. The text-only tokens are used only in Stages~3 and~4. The number of audio tokens is computed as $17.5 \cdot D_s$, where $D_s$ is the audio duration in seconds. Note that an audio clip may be reused across multiple tasks when applicable.}
\label{tab:train_recipe}
\setlength{\tabcolsep}{5pt}
\renewcommand{\arraystretch}{1.15}

\newcolumntype{Y}{>{\raggedright\arraybackslash}X}

\begin{tabularx}{\textwidth}{c p{2.3cm} Y c c c c p{3.2cm}}
\toprule
\textbf{Stage} & \textbf{Goal} & \textbf{Task} & \textbf{Audio Data} & \textbf{Steps} & \textbf{LR} & \textbf{Ctx} & \textbf{Trainable} \\
\midrule
1 &
\makecell[l]{Understanding\\warm-up} &
\makecell[l]{ ASR, Audio Caption \\ Music Caption \\ Lyric recognization} &
3B & 50k & 2e-4 & 1024 &
\makecell[l]{Understanding experts} \\
\midrule
2 &
\makecell[l]{Generation\\warm-up} &
\makecell[l]{TTS, text-to-audio, \\ song/music generation} &
3B & 50k & 2e-4 & 1024 &
\makecell[l]{Generation experts\\+ audio local decoder} \\
\midrule
3 &
\makecell[l]{Audio--text\\pre-training} &
\makecell[l]{understanding and generation tasks, \\ text-only, audio-only} &
50B & 500k & 2e-4 & {1024} &
\makecell[l]{All parameters} \\
\midrule
4 &
\makecell[l]{Mid-training} &
\makecell[l]{subset of stage 3 mixture, \\ new auditory sentence data} &
20B & 300k & 1e-4 & {2048} &
\makecell[l]{All parameters} \\
\bottomrule
\end{tabularx}
\end{table*}

As Figure \ref{fig:overview-training-stage} shows, the training process of UniAudio 2.0 includes four stages. We list the configuration of the four training stages in Table \ref{tab:train_recipe}. For all four stages, we use 64 NVIDIA H100 GPUs to train the model. Below, we describe each stage in detail.

\textbf{Stage~1: Audio understanding warm-up} 
In the first stage, we focus on initializing the audio understanding experts. We train the model using a subset of audio understanding tasks, while freezing all other components. 
To encourage the understanding experts to encode rich semantic information, we introduce an auxiliary semantic distillation objective. Following ReasoningCodec (Section~\ref{sub_sec:semantic_branch}), a lightweight decoder is attached to reconstruct continuous semantic features extracted from frozen WavLM and music SSL encoders. The overall objective includes a reconstruction loss and an LM text loss. After training, we discard the decoder. Formally, the training objective in this stage is defined as
\begin{equation}
\mathcal{L}_{\text{stage1}}
=
\mathcal{L}_{\text{LM}}
+
\lambda_{\text{rec}} \mathcal{L}_{\text{rec}},
\end{equation}
where $\mathcal{L}_{\text{LM}}$ denotes the language modeling loss on text outputs,
and $\mathcal{L}_{\text{rec}}$ is the reconstruction loss for distilling continuous semantic features.
Specifically, the reconstruction loss is given by
\begin{equation}
\mathcal{L}_{\text{rec}}
=
\left\|
D(h) - z_{\text{SSL}}
\right\|_2^2,
\end{equation}

\textbf{Stage~2: Audio generation warm-up.}
In this stage, we update the audio generation expert and the local audio decoder.
We train the model on a subset of audio generation tasks while keeping the understanding and cross-modal experts fixed.
The training objective is defined as
\begin{equation}
\mathcal{L}_{\text{stage2}}
=
\mathcal{L}_{\text{AR}},
\end{equation}
where $\mathcal{L}_{\text{AR}}$ denotes the weighted autoregressive prediction loss over multi-stream audio tokens.

\paragraph{Weighted autoregressive loss over multi-stream tokens.}
Our ReasoningCodec produces $L{=}8$ token streams per audio frame.
Let $s_t^{(\ell)}$ denote the token from stream $\ell\in\{1,\ldots,L\}$ at time step $t\in\{1,\ldots,T\}$.
We define the stream-weighted autoregressive loss as
\begin{equation}
\mathcal{L}_{\text{AR}}
=
-\sum_{t=1}^{T}\sum_{\ell=1}^{L} w_\ell\,
\log p_{\theta}\!\left(s_t^{(\ell)} \mid x,\, s_{<t}^{(1:L)}\right),
\label{eq:weighted_ar_simple}
\end{equation}
where $x$ denotes the conditioning input (e.g., a text prompt), $s_{<t}^{(1:L)}$ denotes all token streams before time step $t$.
In our implementation, we set
\begin{equation}
\mathbf{w}=\Big[\tfrac{2}{8},\,\tfrac{2}{8},\,\tfrac{2}{8},\,\tfrac{1}{8},\,\tfrac{1}{8},\,\tfrac{1}{8},\,\tfrac{1}{8},\,\tfrac{1}{8}\Big].
\end{equation}

\textbf{Stage~3: Audio-text pre-training.}
We jointly update all model parameters on a mixture of audio understanding tasks, audio generation tasks, text-only data, and audio-only data.
This stage aligns the two modalities under a unified autoregressive objective.
We use a maximum context length of 1024 tokens in this stage.
The overall training loss is
\begin{equation}
\mathcal{L}_{\mathrm{AR}}
= \lambda_{\mathrm{text}} \mathcal{L}_{\mathrm{text}}
+ \lambda_{\mathrm{audio}}\, \mathcal{L}_{\mathrm{audio}},
\end{equation}
where $\lambda_{\mathrm{text}}=1.6$ and $\lambda_{\mathrm{audio}}=1$.
Here, $\mathcal{L}_{\mathrm{audio}}$ is instantiated as the weighted token-level loss in Eq.~\eqref{eq:weighted_ar_simple}.
This setting is based on early experiments, which helps preserve the LLM's text capability and improves audio understanding performance.

\textbf{Stage~4: Audio-text mid-training}
In the final stage, we aim to extend the effective context length and enhance generalization to unseen tasks.
We continue training on a subset of the Stage~3 pre-training data, augmented with our constructed auditory sentence data.
This stage encourages the model to model longer and more complex audio-text sequences and improves robustness across diverse task settings. We use a maximum context length of 2048 tokens in this stage. The training objective remains identical to that in Stage~3

\subsection{The details of evaluation data and evaluation metrics} \label{appendix:uniaudio2_eval_data}
UniAudio 2.0 is a multi-task audio foundation model, which supports multiple audio-related tasks. For each task, we follow the commonly used evaluation benchmark and metrics. In the following, we present the details for each task.

\subsubsection{Seen tasks}
\paragraph{ASR} For the ASR task, we follow previous works \cite{qwen-omni,mimo-audio,wenetspeech-yue}, and choose LibriSpeech-test-clean, LibriSpeech-test-other, SEED-TTS-Eval-EN, SEED-TTS-Eval-ZH, and WSYue-ASR-eval \cite{wenetspeech-yue} as the evaluation benchmark. For English datasets, we use WER as the metric. For Chinese and Cantonese, we use CER as the metric. 

\paragraph{TTS} For the TTS task, we use LibriSpeech-test-clean, SEED-TTS-Eval-EN, SEED-TTS-Eval-ZH, WSYue-TTS-eval \cite{wenetspeech-yue} as the benchmark. For evaluation, we use Whisper-large-v3 to evaluate the English speech performance, Paraformer-zh is used for Chinese speech, and SenseVoice-s-Yue is used for Cantonese speech. Furthermore, we also use DNS-MOS to evaluate the speech quality.

\paragraph{Instruct TTS} For the Instruct TTS task, we follow the setting of CapSpeech \cite{capspeech} and use WER, style accuracy, and UTMOSv2 as evaluation metrics. Although our model is not trained on Chinese Instruct TTS, we show that it performs well on Chinese instruction TTS by following InstructTTS \cite{yang2023instructtts} and using the same evaluation dataset. 

\paragraph{Audio Caption} For the audio caption task, we follow Audio Flamingo 3\cite{Audio_flamingo} use CIDER as the evaluation metric. Furthermore, we also use GPT-score to evaluate the model's prediction. The prompt as Box \ref{box:audio_caption_prompt} shows. 

\paragraph{Music Caption} Similarly, for the music caption task, we use MusicCaps test set. The same evaluation metrics are used as the audio caption task.

\paragraph{Audio Generation} For the text-to-audio task, we follow the setting of Stable Audio Open \cite{evans2025stable}, using FD, KL, and CLAP score as the metric. 

\paragraph{Music Generation} For the text-to-music task, we follow MusicGen\cite{musicgen}, using FAD, KL, and CLAP-score as the metrics. 

\paragraph{Song Generation} For the song generation task, we follow SongGen \cite{liu2025songgen}, using their benchmark and evaluation metrics: WER and AudioBox Score \cite{tjandra2025meta}. 

\paragraph{Lyric Recognition} For the lyric recognition task, we use the benchmark from SongGen. 

\subsubsection{Few-shot tasks}
\paragraph{few-shot speech denoising task} We build the few-shot speech denoising evaluation set based on LibriTTS-test-clean and WHAM noise \cite{wichern2019wham} to build the mixture-clean pairs. For the evaluation, we follow Mimo-Audio\cite{mimo-audio}, using PESQ, STOI, WER, DNS-MOS as the metrics. 

\paragraph{Few-shot voice conversion} We build the few-shot voice conversion evaluation set based on VCTK \cite{vctk} dataset. For the evaluation, we follow Mimo-Audio, using WER, Speaker Similarity (SIM), and DNS-MOS as the metrics. 

\paragraph{Few-shot emotion classification} We build the few-shot emotion classification evaluation set based ESD \cite{esd} dataset. We use both English and Chinese splits to build the evaluation set. 

\paragraph{Few-shot sound event classification} We build the few-shot sound event classification evaluation set based on TUT acoustic scenes 2017 \cite{tut2017}. 

\subsection{Zero-shot tasks}
In this study, we define a zero-shot task as one that is never seen during training, and we do not provide any demonstrations at inference time. The model is asked to predict the output based only on the task input. 
\paragraph{text understanding} We follow the standard zero-shot evaluation setting for text LLMs and use the MMLU dataset \cite{mmlu} as the benchmark. We note that some previous work \cite{opuslm} uses a few-shot setting to evaluate text understanding ability; in this study, we directly evaluate its zero-shot ability. 

\paragraph{speech-to-speech/text question answer} During our training stage, we do not add the speech conversation data. Instead, we build a lot of auditory sentence without explicit instruction. Thus, we view the speech-to-speech/text question answer as the zero-shot task. We follow LLAMA-Omni \cite{llama-omni}, and use the InstructS2S-Eval as the benchmark. Following LLAMA-Omni, GPT-score is used as the metric. We use the same prompt to evaluate the performance as LLAMA-Omni.

\paragraph{Dysarthric Speech Recognition} Dysarthric speech recognition is very similar to the ASR task, but it asks the model to recognize dysarthric speech. Each utterance only includes one word. We also include it as a zero-shot setting, since the model was not trained on such data. 

\paragraph{Audio prompt and caption guided TTS} We define a new task: using audio prompt to provide the timbre and the caption to guide the speaking style. We use speaker similarity, style accuracy, WER, UTMOSv2 as the metrics.

\paragraph{speech-sound generation} We define a new task: asking the model to generate speech and corresponding sound event. The input includes sound event tag and speech content. We use WER, CLAP score, and UTMOSv2 as the metrics. 


\begin{tcolorbox}[
  enhanced,
  colback=blue!3,
  colframe=blue!55,
  boxrule=0.6pt,
  arc=2mm,
  left=6pt,right=6pt,top=6pt,bottom=6pt,
  title=\textbf{Prompt: Audio Caption Evaluation},
  fonttitle=\bfseries,
]
\label{box:audio_caption_prompt}
\small
\ttfamily
\begin{lstlisting}[basicstyle=\ttfamily\small,breaklines=true]
You are an expert evaluator for audio captioning.

Given:
- PRED (model caption)
- GT (human reference caption)

Evaluate PRED against GT and output ONLY a valid JSON object.

Scoring (integers 0-10):
- relevance: whether PRED describes the same sound event(s) as GT
- fluency: whether PRED covers key details without grammaticality
Also provide:
- overall: integer 0-10 (holistic score)
- brief_reason: <= 60 words, concise explanation focusing on mismatches
\end{lstlisting}
\end{tcolorbox}

\subsection{The details of experimental results} \label{appendix:uniaudio_detailed_exp}
Due to the page limit, we only choose some representative works in Table \ref{tab:seen_tasks}. In this part, we will compare with more baseline models on different tasks. 

\subsubsection{ASR performance comparison}

\begin{table}[t]
\centering
\caption{ASR performance on LibriSpeech and Seed-TTS benchmarks. Note that we do not find a useful ASR instruction for Step-Audio-chat, thus we directly use the official reported results on Librispeech benchmark.}
\label{tab:appendix_asr_librispeech_seedtts}
\setlength{\tabcolsep}{7pt}
\renewcommand{\arraystretch}{1.15}
\begin{tabular}{lcccc}
\toprule
\textbf{Model} & \textbf{LibriSpeech-clean} & \textbf{Libri-other} & \textbf{Seed-TTS ZH} & \textbf{Seed-TTS EN} \\
\midrule
Mimo-Audio-Instruct-7B \cite{mimo-audio}        & 3.50    & 35.43 & 29.81 & 7.01 \\
Qwen2.5-Omni-7B \cite{qwen-omni}     & 3.92  & 5.52  & 1.3 & 2.89 \\
Step-Audio-chat-3B \cite{stepaudio}   & 3.11    & 8.44    & --    & --    \\
Whisper-large-v3 \cite{whisper}  & 1.81  & 3.55  & 6.8   & 1.47 \\
UniAudio 2.0 (Ours)              & 2.71  & 6.33  & 2.6 & 2.14 \\
\bottomrule
\end{tabular}
\end{table}

\begin{table}[t]
\centering
\caption{ASR results on WenetSpeech-Yue ASR Benchmark.}
\label{tab:yue_asr}
\setlength{\tabcolsep}{8pt}
\renewcommand{\arraystretch}{1.15}
\begin{tabular}{ccc}
\toprule
\textbf{Model} & \textbf{Long sentence} & \textbf{Short sentence} \\
\midrule
Whisper-large-v3 \cite{whisper} & 36.8 & 31.5 \\
SenseVoice-Yue \cite{wenetspeech-yue}   & 15.7 & 6.0  \\
Qwen2.5-Omni-7B \cite{qwen-omni}        & 23.5 & 31.0 \\
UniAudio 2.0 (Ours)             & 12.1 & 7.7  \\
\bottomrule
\end{tabular}
\end{table}

Table \ref{tab:appendix_asr_librispeech_seedtts} shows the ASR performance comparison on Chinese and English benchmarks. Table \ref{tab:yue_asr} shows the Cantonese ASR performance. We compare with SenseVoice-Yue \cite{wenetspeech-yue}, Whisper-large-v3 \cite{whisper}, and Qwen2.5-Omni. 

\subsubsection{TTS performance comparison}

\begin{table}[t]
\centering
\caption{TTS results on Seed-TTS (EN/ZH), LibriSpeech-clean, and Cantonese TTS. We report the WER and DNS-MOS score (WER / DNS-MOS). The CosyVoice-Yue is the official checkpoint from WenetSpeech-Yue \cite{wenetspeech-yue}.}
\label{tab:appendix_exp_tts}
\setlength{\tabcolsep}{8pt}
\renewcommand{\arraystretch}{1.15}
\begin{tabular}{ccccc}
\toprule
\textbf{Model} & \textbf{Seed-TTS-EN} & \textbf{Seed-TTS-ZH} & \textbf{LibriSpeech-clean} & \textbf{WenetSpeech-Yue-TTS}\\
\midrule
MiMo-Audio-In-7B & 4.74 / 3.27 & 1.93 / 3.34 & 5.30 / 3.35 & - \\
Qwen2.5-Omni-7B    & \textbf{3.10} / 3.62 & \textbf{1.21} / 3.65 & 4.28 / 3.72  & \textbf{12.2} / 3.36 \\
CosyVoice-Yue    & - & - & -  & 13.9 / 3.34 \\
UniAudio 2.0 (Ours)            & 3.63 / \textbf{3.80} & 2.30 / \textbf{3.82} & \textbf{3.46} / \textbf{3.88}  & 12.5 / \textbf{3.41} \\
\bottomrule
\end{tabular}
\end{table}

Table \ref{tab:appendix_exp_tts} shows the TTS performance on Seed-TTS-eval (EN/ZH), LibriSpeech-clean (EN), and WenetSpeech-Yue-TTS (Cantonese) benchmarks. 

\begin{table*}[t]
\centering
\caption{Ablation studies for multi-stage training, We include additional tasks that are not reported in Table~\ref{tab:ablation_stage}. For Few-shot VC and Few-shot Sound, we report results under the 1-shot setting. We use \texttt{NA} to indicate that the model cannot perform these few-shot task without the stage 4 training.}
\label{tab:appendix_ablation_more}
\setlength{\tabcolsep}{6pt}
\renewcommand{\arraystretch}{1.15}
\begin{tabular}{c c c c c c c}
\toprule
\textbf{Setting} & \textbf{AudioCaps} & \textbf{MusicCaps} & \textbf{Lyric Recognition} & \textbf{InstructTTS} & \textbf{Few-shot VC} & \textbf{Few-shot Sound} \\
\midrule
w/o Stage 4
& 0.34
& \makecell[c]{4.56}
& \makecell[c]{29.7}
& \makecell[c]{8.8/35.2/3.41}
& \makecell[c]{NA}
& \makecell[c]{NA}
\\
\midrule
w/o Experts
& 0.22
& \makecell[c]{4.33}
& \makecell[c]{34.3}
& \makecell[c]{9.7/22.9/3.21}
& \makecell[c]{NA}
& \makecell[c]{NA}
\\
\midrule
1B
& 0.30
& \makecell[c]{4.52}
& \makecell[c]{32.8}
& \makecell[c]{7.9/31.1/3.16}
& \makecell[c]{21.3/0.82/3.66}
& \makecell[c]{48.7}
\\
\midrule
Ours (3B)
& 0.69
& \makecell[c]{5.14}
& \makecell[c]{28.57}
& \makecell[c]{7.3/42.3/3.38}
& \makecell[c]{18.61/0.89/3.74}
& \makecell[c]{59.8}
\\
\bottomrule
\end{tabular}
\end{table*}

\subsubsection{Audio/music caption performance comparison}

\begin{table}[t]
\centering
\caption{Captioning results on audio and music.}
\label{tab:appendix_caption_audio_music}
\setlength{\tabcolsep}{6pt}
\renewcommand{\arraystretch}{1.15}
\begin{tabular}{cccccc}
\toprule
& \multicolumn{3}{c}{\textbf{Audio Caption}} & \multicolumn{2}{c}{\textbf{Music Caption}} \\
\cmidrule(lr){2-4}\cmidrule(lr){5-6}
\textbf{Model} & \textbf{CIDEr} & \textbf{Relevance} & \textbf{Fluency} & \textbf{Relevance} & \textbf{Fluency} \\
\midrule
SALMON 13B \cite{salmonn}         & 0.355 & 5.41 & 8.06 & 3.24 & 7.92 \\
Music Flamingo \cite{musicflamingo}    & --    & --   & --   & 4.74 & 8.19 \\
Audio Flamingo \cite{flamingo1}     & 0.51  & 5.25 & 8.01 & --   & --   \\
Audio Flamingo 2 \cite{flamingo2}   & 0.58  & 5.71 & 8.13 & --   & --   \\
Audio Flamingo 3 \cite{Audio_flamingo}  & 0.79 & 6.34 & 8.24 & 5.77 & 8.24 \\
Qwen2.5-Omni-7B \cite{qwen-omni}       & 0.39  & 5.61 & 8.21 & 5.33 & 7.90 \\
UniAudio 2.0 (Ours)               & 0.69  & 5.51 & 8.31 & 5.14 & 8.00 \\
\bottomrule
\end{tabular}
\end{table}

Table \ref{tab:appendix_caption_audio_music} shows the performance comparison on audio and music caption benchmarks. 

\subsubsection{Audio generation performance comparison}

\begin{table}[t]
\centering
\caption{Audio generation results. We report these baseline's results from Stable Audio Open paper.}
\label{tab:appendix_audio_generation}
\setlength{\tabcolsep}{8pt}
\renewcommand{\arraystretch}{1.15}
\begin{tabular}{cccc}
\toprule
\textbf{Model} & \textbf{KL} & \textbf{FD} & \textbf{CLAP-score} \\
\midrule
AudioLDM2-large \cite{liu2023audioldm}  & 1.57 & 170.31 & 0.41 \\
Stable Audio Open \cite{evans2025stable} & 2.14 & 78.24  & 0.29 \\
AudioGen \cite{kreuk2022audiogen}        & 1.42 & 186.53 & 0.45 \\
UniAudio 2.0 (Ours)             & 3.26 & 50.69  & 0.17 \\
\bottomrule
\end{tabular}
\end{table}


Table \ref{tab:appendix_audio_generation} shows the text-to-audio generation performance on audio caption tasks. 


\subsubsection{Zero-shot experiments for 1B model} \label{appendix:zero-shot_1b_exp}
\begin{table}[t]
\centering
\caption{Zero-shot results comparison on different tasks. Metrics: MMLU \cite{mmlu} reports Acc (\%); InstructS2S-Eval reports S2S/S2T Acc; DSR reports WER (\%); A-I-TTS reports SIM / Style-Acc (\%) / WER (\%) / UTMOSv2; Speech$\rightarrow$Sound reports WER (\%) / CLAP-score / UTMOSv2. S2S denotes speech-to-speech instruction-following and S2T denotes speech-to-text instruction-following. A-I-TTS denotes audio+caption guided speech generation.}
\label{tab:appendix_zero_shot_1B}
\setlength{\tabcolsep}{6pt}
\renewcommand{\arraystretch}{1.12}
\begin{tabular}{c c c}
\toprule
\textbf{Task} & \textbf{Model} & \textbf{Score} \\
\midrule

\multirow{3}{*}{Text}
& LLAMA 3.2 1B \cite{llama3} & 34.14 \\
& LLAMA 3.2 3B  \cite{llama3} & 47.63 \\
& UniAudio 2.0 (Ours 3B) & 44.1 \\
& Ablation (Ours 1B)    & 30.2 \\
\midrule

\multirow{3}{*}{S2S}
& LLAMA-Omni \cite{llama-omni} & 3.47 / 3.99 \\
& SpeechGPT \cite{speechgpt}& 2.19 / 2.98 \\
& Ours       & 2.16 / 3.66 \\
& Ablation (Ours 1B)    & 1.12 / 1.41 \\
\midrule

\multirow{2}{*}{DSR}
& Qwen2.5-Omni-7B \cite{qwen-omni} & 80.6 \\
& UniAudio 2.0 (Ours 3B)    & 19.4 \\
& Ablation (Ours 1B)    & 61.1 \\
\midrule

A-I-TTS
& UniAudio 2.0 (Ours 3B) & 0.89 / 32.62 / 11.57 / 2.87 \\
& Ablation (Ours 1B) & 0.62 / 5.2 / 14.8 / 2.43 \\
\midrule

Speech-S
& UniAudio 2.0 (Ours 3B) & 6.15 / 0.11 / 2.96 \\
& Ablation (Ours 1B) & 6.79 / 0.04 / 2.82 \\
\bottomrule
\end{tabular}
\end{table}

Table \ref{tab:appendix_zero_shot_1B} shows the performance comparison between our 3B and 1B models. We find that the 1B model's zero-shot performance is far behind the 3B version, which further highlights the importance of model size for model generalization. In Table \ref{tab:appendix_ablation_more}, we report more task performance comparison about the ablation study.

\section{Limitation} \label{appendix:limitation}

In this study, we focus on building a multi-task audio foundation model that supports diverse audio understanding and generation tasks. It can also generalize to many unseen tasks in few-shot or zero-shot settings. However, several limitations remain.

(1) To improve reconstruction quality for sound and music, we adopt a flow-based decoder to recover waveforms from semantic tokens. The multi-step decoding procedure in flow matching increases inference latency for generation. In the future, it is necessary to explore few-step decoding (e.g., two steps) to better balance quality and generation speed.

(2) Although UniAudio 2.0 demonstrates the ability to handle unseen tasks, there is still room for improvement. In addition, it has not yet been shown to solve arbitrary audio-related tasks. We acknowledge that the set of supported unseen tasks is closely related to the training data. For example, the model currently cannot handle speech diarization, likely because we do not include diarization- or duration-related supervision during training.  

(3) Due to limited GPU resources, we have not fully explored scaling behaviors (i.e., scaling laws) of UniAudio 2.0. We only conduct experiments on 1B- and 3B-parameter variants. In the future, scaling to 7B and larger models is a promising direction.

(4) Due to the relatively limited amount of sound and music data compared to speech data, UniAudio~2.0 currently performs better on speech-related tasks. In future work, expanding and improving sound and music datasets is expected to further enhance performance in these domains.

(5) This work primarily focuses on pre-training design choices, such as the audio tokenizer and the unified LLM architecture. As a result, we do not extensively investigate post-training strategies (e.g., multi-task SFT and reinforcement learning). We plan to incorporate more post-training techniques to further improve UniAudio 2.0.

(6) We acknowledge that the set of compared models is not exhaustive. This is partly because many related models are not publicly available, and partly because our framework supports a broad spectrum of tasks, which makes comprehensive comparisons challenging. We respect and appreciate all prior work in this area, even if some are not explicitly discussed due to space limitations. We also do not claim that UniAudio~2.0 universally outperforms all existing approaches; instead, different model architectures, different special task design (e.g. special models for TTS, ASR, diffusion-based unified models) also offer complementary strengths.

\end{document}